\documentclass[showpacs,aps,prd,reprint,groupedaddress,superscriptaddress,preprintnumbers,floatfix,a4paper,nofootinbib,amsmath,amssymb]{revtex4-1}
\usepackage{color}
\usepackage{amsmath}
\usepackage{footmisc}
\usepackage{amssymb}
\usepackage{dcolumn}  
\usepackage{slashed}
\usepackage{isotope}
\usepackage{graphicx}
\usepackage{mathtools}
\usepackage[utf8]{inputenc}
\usepackage{multirow}
\usepackage[shortlabels]{enumitem}
\usepackage{hyperref}
\usepackage{soul}
\usepackage{dutchcal}
\hypersetup{colorlinks, linkcolor = [rgb]{0,0.0,0.75}, citecolor = [rgb]{0,0.0,0.75}, urlcolor = [rgb]{0,0.0,0.75}}

\makeatletter
\g@addto@macro\bfseries{\boldmath}
\makeatother

\begin{document}
\allowdisplaybreaks

\title{Study of three-flavored heavy dibaryons using lattice QCD}

\author{Parikshit M. Junnarkar}
\email{parikshit@theorie.ikp.physik.tu-darmstadt.de}
\affiliation{Institut f{\"u}r  Kernphysik, Technische Universit{\"a}t Darmstadt,\\ Schlossgartenstra{\ss}e 2, Darmstadt 64289, Germany.}

\author{Nilmani Mathur}
\email{nilmani@theory.tifr.res.in}
\affiliation{Department of Theoretical Physics, \\
Tata Institute of Fundamental Research,
1 Homi Bhabha Road, Mumbai 400005, India.}




\begin{abstract}
We present results of the first lattice QCD calculation of three-flavored heavy dibaryons both in the flavor-symmetric and in the
antisymmetric channels.  These dibaryons have spin zero, and the
three-flavored states are constructed using various possible combinations of  quark flavors with at least one of them as the charm ($c$) or the bottom ($b$) quark {\it i.e.}, namely, $\mathcal{H}_c(cudcud), \mathcal{H}_b(budbud), \mathcal{H}_{bcs}(bcsbcs)$, $\mathcal{H}_{csl}(cslcsl), \mathcal{H}_{bsl}(bslbsl)$ and $\mathcal{H}_{bcl}(bclbcl)$; $ l \in u,d$. 
We compute the ground state masses of these dibaryons and the calculations
are performed on three $N_f = 2 + 1 + 1$ HISQ gauge ensembles of the
MILC collaboration, with lattice spacings $a = 0.1207, 0.0888 $ and
$0.0582 $ fm. A relativistic overlap action is employed for the
valence light and charm quarks and a non-relativistic-QCD Hamiltonian with improved coefficients is used for the bottom quarks.  Unlike the doubly heavy tetraquarks, deuteron-like heavy dibaryons and dibaryons with only bottom quarks, for which lattice QCD calculations have predicted deeply bound strong-interactions-stable states, for these $\mathcal{H}_c(cudcud), \mathcal{H}_b(budbud), \mathcal{H}_{csl}(cslcsl), \mathcal{H}_{bsl}(bslbsl)$ dibaryons we do not find any such deeply bound state at the physical quark masses. However, for $\mathcal{H}_{bcs}(bcsbcs)$, our results indicate the presence of an energy level $29\pm 24$ MeV below the lowest two-baryon elastic threshold, which could be relevant for its future experimental searches. Moreover, we find that the energy difference between the lowest state and the lowest elastic threshold, which could well be interpreted as the binding energy for such heavy dibaryons ($\mathcal{H}_{bcl}$), increases with the increase of quark masses ($m_l > m_s$). Taken together, our findings indicate the possibility of the existence of the physical $\mathcal{H}_{bcs}$ dibaryon while all other physical three-flavored dibaryons are much closer to their thresholds suggesting either they are weakly bound or unbound, resolving which requires further detail study. Our results also point that the binding of a dibaryon configuration becomes stronger with the increase of its valence quark masses which suggests an interesting aspect of strong interaction dynamics at multiple scales.
\end{abstract}

\pacs{12.38.Gc, 12.38.-t, 14.20.Mr, 14.40.Pq 
   } 

\keywords{multi-baryon spectroscopy, heavy dibaryons, lattice QCD}

\maketitle


\section{\label{sec:intro}Introduction}
Quantum chromodynamics (QCD), the theory of strong interactions of quarks and gluons, predicts a very rich energy spectra of hadronic states comprising various quark flavors from light to the bottom \cite{Zyla:2020zbs}.  While most of the observed  hadrons are classified as mesons and baryons within quark models, QCD also allows the existence of other bound state configurations of quarks (and antiquarks), which are generically known as {\it exotic hadrons}. Indeed  the recent discovery of a large number of new subatomic particles, the so-called X,Y Z hadrons \cite{Esposito:2014rxa, Ali:2017jda, Olsen:2017bmm, Guo:2017jvc, Karliner:2017qhf, Brambilla:2019esw,Liu:2019zoy}, including tetraquarks \cite{Belle:2003nnu, Belle:2011aa, Aaij:2013zoa, Aaij:2014jqa, BESIII:2020qkh, LHCb:2021vvq} and pentaquarks, \cite{Aaij:2015tga, LHCb:2019kea} have confirmed the existence of a new class of subatomic particles in Nature. These discoveries naturally have created tremendous excitement to the field of hadron spectroscopy \cite{Zyla:2020zbs, Esposito:2014rxa, Ali:2017jda, Olsen:2017bmm, Guo:2017jvc, Karliner:2017qhf, Brambilla:2019esw,Liu:2019zoy}. In terms of the number of valence quark content, these recent discoveries have so far been limited to four (tetra)- and five (penta)-quark states, and except the possible finding of a broad  $d^*(2380)$ resonance \cite{PhysRevLett.112.202301}\footnote{This resonance peak structure is also reported to be tied to a triangle singularity \cite{Molina:2021bwp}} no new six (hexa)-quark state has yet been discovered. It is therefore natural to investigate the existence of hadrons with six valence quark configurations within QCD, with the goal that QCD-predicted such states can guide in discovering them in future at high energy laboratories.

In our visible universe, so far deuteron is found to be the only {\it stable} six-quark bound state with a binding energy of about 2.2 MeV which has been modelled to be the result of a many body interactions of two nucleons \cite{KAPLAN1998329,RevModPhys.81.1773}.
The so-called $H(udsuds)$-dibaryon is another highly speculated compact six-quark bound state with strangeness $S=-2$, spin $J = 0$ and isospin $I = 0$. In the first calculation of $H$-dibaryon, a binding of 70 MeV was predicted using the MIT bag model~\cite{Jaffe:1976yi}. Thereafter this state had seen a thorough investigation in the past four decades through various model studies \cite{Donoghue:1986zd, Golowich:1992zw, Haidenbauer:2011ah, Haidenbauer:2015zqb}. 
However, till date, experimental searches have ruled out the existence of such a deeply bound state~\cite{Takahashi:2001nm,Kim:2013vym}. The $H$ dibaryon has also been a subject of numerous lattice QCD calculations in recent years and the results have also ruled out a deeply bound state, and rather indicate an unbound or a very weakly bound state if at all it exists \cite{NPLQCD:2010ocs,Inoue:2010es,Luo:2011ar,Francis:2018qch,Green:2021qol}.
 However, to confirm the existence of such a state through lattice QCD calculations, it is 
essential to perform chiral and continuum extrapolations to the physical limits, along with a detailed 
finite-volume amplitude analysis of the pole distribution in the scattering amplitude across the complex energy plane \cite{Luscher:1990ck}.
Light dibaryons have also been studied in other channels. For example, in the $SU(3)_f$
quark model, one expects attraction in dibaryon states without quark
Pauli blocking and with attractive color-spin interaction from
one-gluon exchange, $\Lambda\Lambda-\Sigma\Sigma-N\Xi$ (H-channel),
$\Delta\Delta$, and $N\Omega$ states \cite{PhysRevD.38.298,Gal:2015rev}. Recent experimental and
theoretical studies based on femtoscopy also suggest the possibility of a
$N\Omega$ bound state in the $J^\pi=2^+$ channel \cite{STAR:2018uho,ALICE:2020mfd,PhysRevC.101.015201}.

In the spirit of finding $H$-dibaryon, six-quark states in the isosinglet channel but with heavy quarks have also been explored in several model calculations with results not favoring a bound state \cite{Meguro:2011nr, Vijande:2016nzk, Richard:2016eis, Dong:2021bvy}. On the contrary, recently in a first principles lattice QCD calculation it was found that strong-interactions-stable deuteron-like heavy dibaryons can exist if at least two of the quarks in a dibaryon have heavy flavors \cite{Junnarkar:2019equ}. Lattice calculations have also been performed recently for single-flavored heavy dibaryons with charm  as well as bottom flavors. While using HALQCD potential method the bindings for $^1S_0$ $\Omega_{ccc}$-$\Omega_{ccc}$ has been reported to be about $ -5$ MeV \cite{Lyu:2021qsh}, a direct calculation for $\Omega_{bbb}$-$\Omega_{bbb}$ dibaryons finds a deep binding of about $-90$ MeV in the $^1S_0$ channel \cite{Mathur:2022nez}.

Hence it is interesting to investigate the nature of quark mass dependence of $H$-like three-flavored dibaryons, particularly whether such a state is bound with one or more heavy quark content. Any clear indication on the existence of heavy $H$-like three-flavored dibaryon from lattice QCD calculations will be attractive for searching them at high energy laboratories. Moreover, a detailed study of quark mass dependence on the binding energy can further reveal the intriguing dynamics of heavy quarks in dibaryons which can further illuminate our knowledge of strong interactions at multiple scales.

Lattice QCD is an ideal tool for studying  multi-hadron systems since in addition to being a first principles method, it is also possible to obtain quantitative results through lattice QCD at any quark masses, including at their unphysical values. As a result it provides a unique tool for systematic study of the quark mass dependence of binding energies, which otherwise is not possible to obtain. Moreover, with adequate computing resources it is possible to keep track of all the systematic uncertainties associated with such calculations.
In recent years, beside the regular single-hadron energy spectra, lattice QCD methodology has been used successfully in studying multi-mesons as well as multi-baryons and nuclear systems \cite{BEANE20111,PhysRevD.85.054511,PhysRevD.73.054503,Ishii:2006ec,Aoki:2012tk,PhysRevD.87.034505,PhysRevLett.113.182001,PhysRevLett.118.022002}. These studies involve multifold challenges, namely evaluating a large number of Wick contractions, addressing the poor signal-to-noise ratio and the associated finite volume effects.  
Moreover, because of their exceedingly large numbers, the computational cost of Wick contractions for multi-baryon states can even exceed the cost of quark propagator computation and new algorithms are necessary to address this issue. New methods have indeed been developed, namely sink momentum projection~\cite{Detmold:2012eu}, evaluating simultaneous contractions~\cite{Doi:2012xd} and manipulation of permutation symmetry through tensor properties \cite{Humphrey:2022yjc}, which help to somewhat mitigate the computational cost.  
Besides Wick contractions, the issue of reliable ground state determination is also important due to the worsening of signal-to-noise ratio for multi-hadron systems \cite{BEANE20111}. 
The main challenge here is the need for a variational calculation with multi-baryon operators with good sources which can clearly separate the ground state from excited states.
To overcome this issue a study has recently been performed employing both methods of point sources and distillation, and it reaches to the conclusion that more precise results for multi-hadron systems can indeed be obtained through operator smearing through distillation~\cite{Francis:2018qch}.  Another crucial challenge in searching for bound states is a detailed finite-volume amplitude analysis of the pole distribution in the scattering amplitude across the complex energy plane \cite{Luscher:1990ck}.
This requires calculations either at different physical volumes or in different momentum frames. Both of these require significant computational resources. 

Working with heavy quarks can somewhat mitigate some of the aforementioned challenges associated with multi-baryon systems. For example, it is expected that the effect of chiral dynamics will be less severe in heavy multi-baryon systems and hence the signal-to-noise ratio in correlation functions may possibly be improved. Moreover, due to the presence of heavy hadrons at thresholds, a relatively large suppression of the finite volume effects on the extracted energy levels is expected for heavy dibaryons. Of course, light quarks in heavy dibaryons can still produce not-so-good  signal-to-noise ratio and the presence of a light baryon at the threshold can also enhance the finite volume effects. While there have been numerous multi-baryon lattice QCD studies with light and strange quarks, there is almost no investigation on multi-baryons with heavy flavors until recently~\cite{Junnarkar:2019equ, Lyu:2021qsh, Mathur:2022nez}. As mentioned before, in a first of its kind, deuteron-like dibaryons with heavy quarks were investigated recently and it was found that such states with charmed-bottom, strange-bottom and strange-charmed flavor combinations have large binding energies~\cite{Junnarkar:2019equ}. Large binding has also been found for the dibaryon which has only bottom quarks  \cite{Mathur:2022nez}. 
Interestingly, multiple lattice groups have predicted the large binding energies of tetraquarks with heavy quark contents \cite{Bicudo:2012qt,Francis:2016hui,Junnarkar:2018twb,Leskovec:2019ioa}. Further, the exotics states that have been discovered in the last two decades are all have heavy flavors \cite{Zyla:2020zbs, Belle:2003nnu, Belle:2011aa, Aaij:2013zoa, Aaij:2014jqa, BESIII:2020qkh, LHCb:2021vvq,Aaij:2015tga, LHCb:2019kea}. 

Inspired by those theoretical and experimental studies, here we perform a pilot study of three-flavored heavy dibaryons
using lattice QCD and report the findings.
In particular, we investigate the spin-0 $H$-like dibaryons in the isosinglet $(I=0)$ and isotriplet $(I=1)$ channels by replacing the two strange quarks with the heavy flavors yielding dibaryons with the following quark contents: $Qq_1q_2Qq_1q_2$ where $Q \equiv c$ or $b$ and $q_i \equiv u, d, s$.
These states are therefore charm and bottom quark analogues of $H$-dibaryon and also belong to $SU(3)$ 27-plet.
Our results from this study do not support any physical bound state with deep bindings for any three-flavored $H$-like heavy dibaryons for isosinglet configurations. Energy levels for the isotriplet dibaryons are found to be even higher, suggesting possible scattering states. For example, for the cases of $\mathcal{H}_{c}(cudcud), \mathcal{H}_{b}(budbud),$ and $\mathcal{H}_{csl}(cslcsl), \mathcal{H}_{bsl}(bslbsl), \mathcal{H}_{bsl}(bclbsl)$; $l \in u, d$, we do not find any energy level far below their respective lowest threshold. However, for the physical isosinglet $\mathcal{H}_{bcs}(bcsbcs)$ dibaryon we find an energy level consistently below its lowest elastic threshold in all lattice ensembles utilized in this work. When extrapolated to the continuum, this energy difference is found to be $-29(24)$ MeV. Although the large errobar prohibits us to reach to a conclusive evidence for the existence of $\mathcal{H}_{bcs}(bcsbcs)$, this pilot study definitely indicates that if there is any deeply bound three-flavor dibaryon, then it has to be $\mathcal{H}_{bcs}(bcsbcs)$. A further study with a large statistics and more control over systematics is now called-for to conclude about the binding nature of this state, and if a positive result is found from such a study there will be enough motivation for searching this dibaryon experimentally.

The paper is organized as below. In section \ref{sec:latt}, we discuss the details of our lattice set up. Next we elaborate the relevant interpolating fields used in this study. In section \ref{sec:results}, we discuss the details of the analysis and present the results. Finally, in section \ref{sec:conclusion}, we provide a discussion of the results from this study and possible future outlooks.
\section{\label{sec:latt}Lattice set up and Interpolating Operators}
The lattice set up that we employ for this work is similar to the one used in our previous works in Refs.~\cite{PhysRevLett.121.202002, Mathur:2018rwu, Junnarkar:2018twb,Junnarkar:2019equ}, but we describe it here for completeness. The gauge ensembles utilized for this calculation are generated by the MILC collaboration  with $N_f=2+1+1$ flavors of sea quarks using HISQ action~\cite{Bazavov:2012xda}. We employ following three sets of lattice ensembles as listed in Table~\ref{tab:pars}.
\begingroup
\renewcommand*{\arraystretch}{1.5}
\begin{table}[ht]
	\centering
	\begin{tabular}{cccc} \hline \hline 
	$L^3 \times T$     & $m^{\text{sea}}_\pi$ (MeV) & $m_\pi \text{L}$ & $a$ (fm)\\ \hline \hline    
	$24^3 \times 64 $  &              305.3           & 4.54       & 0.1207(11) \\ \hline 
        $32^3 \times 96 $  &              312.7           & 4.50       & 0.0888(8) \\ \hline 
	$48^3 \times 144 $ &              319.3           & 4.51       & 0.0582(5) \\ \hline \hline
	\end{tabular}
	\caption{\label{tab:pars}{Parameters of lattice QCD ensembles used in this work.}}
	\end{table} 
	\endgroup
For these ensembles the strange and charm sea-quark masses were set to their physical values on these ensembles.
The scale was set by the MILC collaboration using the $r_1$ parameter and the corresponding lattice spacings are listed in the last column of Table~\ref{tab:pars}. These values were also found to be consistent with the scales obtained
through Wilson flow~\cite{Bazavov:2015yea}.

For the valence fermions, as in our earlier works~\cite{PhysRevLett.121.202002, Mathur:2018rwu, Junnarkar:2018twb,Junnarkar:2019equ}, we employ a relativistic overlap action for light to charm quarks. A gauge fixed wall source is utilized to compute the valence propagators.
The strange quark mass is tuned by setting the unphysical pseudoscalar mass $s \bar{s}$ to 688 MeV \cite{PhysRevD.91.054508}.
The charm quark  mass is set to its physical value by equating the kinetic mass of the spin-averaged $1S$-charmonia, ${1\over 4}(3M_{J/\psi}+M_{\eta_c})$, to its experimental value.
For the bottom quark, we use a non-relativistic QCD (NRQCD) formulation~\cite{Lepage:1992tx}, where the Hamiltonian includes all the terms up to $1/(am_b)^2$ and leading order term of $1/(am_b)^3$, with $m_b$ as the bare bottom quark mass.
The NRQCD Hamiltonian is given by $H = H_0 + \Delta H$, where  the interaction term, $\Delta H$, as used here, is given by,
\begin{eqnarray}
\Delta H &=& -c_1 \frac{(\Delta^{(2)})^2}{8(am_b)^3} + c_2 \frac{i}{8 (am_b)^3}(\nabla \cdot \tilde{E}  - \tilde{E} \cdot \nabla) \nonumber \\
&& -c_3 \frac{1}{8 (m_b)^2} \sigma \cdot (\nabla \times \tilde{E}  - \tilde{E} \times \nabla) - c_4 \frac{1}{2 am_b} \sigma \cdot \tilde{B} \nonumber \\
&& +  c_5 \frac{\Delta^{(4)}}{24 a m_b} - c_6 \frac{(\Delta^{(2)})^2}{16 (am_b)^2}, \\ \nonumber
\end{eqnarray}
with $c_{1..6}$ as the tuned improvement coefficients \cite{PhysRevD.85.054509}.
The bottom quark mass is tuned by setting the kinetic mass of the spin-averaged
 $1S$-bottomonia, ${1\over 4}(3M_{\Upsilon}+M_{\eta_b})$, to its experimental value. The bottom quark propagators are computed following the usual NRQCD evolution of the above Hamiltonian.

As mentioned above, we use overlap fermions for the light valence quark propagators, and in Table~\ref{tab:mqs} below, we list the range of valence quark masses and the corresponding pseudoscalar meson masses that we use for three different ensembles.
\begingroup
\renewcommand*{\arraystretch}{1.2}
\begin{table}[ht]
	\centering
	\begin{tabular}{ccc}\hline \hline
		$L^3 \times T$  & $a$ (fm) & $m_\pi$ (MeV)  \\ \hline \hline
		$24^3 \times 64$    & 0.1207(14)    &  688 \\ 
	\hline \hline
		$32^3 \times 96$ & 0.0888(5)&  688   \\ 
            \hline \hline                                                                                $48^3 \times 144$ & 0.0582(5)  & 9399\\
                               &            & 6175\\
                               &            & 5146 \\
                               &            & 4120 \\                                                          &            & 2984  \\           
                               &            & 688  \\
			       &            & 645  \\
			       &            & 576  \\
		               &            & 550  \\
		               &            & 480  \\
		               
		 \hline \hline	\end{tabular}
	\caption{\label{tab:mqs}{Range of pseudoscalar meson masses on each of the ensembles used in this work.}}
\end{table} 
\endgroup

We now elaborate the interpolating operators used in calculating the three-flavored heavy dibaryons in this work.
These are local six-quark interpolating operators projected onto the antisymmetric  and symmetric flavor representations and have already been explored in the literature in the context of searches for the $H$ dibaryon in previous lattice calculations~\cite{Wetzorke:1999rt,Wetzorke:2002mx,Junnarkar:2018twb} as well as in model studies~\cite{Golowich:1992zw,Donoghue:1986zd}.
In the context of the present work, the charm and bottom quarks replace the strange quarks and accordingly the quantum numbers of the states.
The interpolating operator is constructed as products of three diquarks as, 
\begin{align}\label{eq:hexaquark}
[\mathbcal{abcdef}] = \epsilon_{ijk} &\epsilon_{lmn} \Big( \mathbcal{b}^i C\gamma_5 P_+ \mathbcal{c}^j \Big) \\ \nonumber
& \times\Big( \mathbcal{e}^l C\gamma_5 P_+ \mathbcal{f}^m \Big) \Big( \mathbcal{a}^k C\gamma_5 P_+ \mathbcal{d}^n \Big) ({\vec{x}}, t)\,,
\end{align}
where the alphabets with bold calligraphy, $(\mathbcal{a},\mathbcal{b}, \mathbcal{c}, \mathbcal{d}, \mathbcal{e}, \mathbcal{f})$, indicate quark fields at the site $({\vec{x}}, t)$, $C$ is the charge conjugation operator and $P_+ \equiv (1 + \gamma_4)$ is used for the positive parity projection.
With this notation, the antisymmetric combination is represented as~\cite{Wetzorke:1999rt},
\begin{align}\label{eq:sixq_0}
\mathcal{H}^{\text{AS}}_{Qql} &= \Big( \big[QlqQlq \big] - \big[lqlQqQ \big] - \big[ qlqQlQ \big]  \Big),
\end{align}
where the flavor $Q\in (c,b)$ represents the heavy quark flavor of charm $c$ or bottom $b$.  Similarly the flavor $q \in (c,s)$ when $Q =  b$,  and the flavor $l$ is understood to be the light flavor.
In addition to the antisymmetric channel, we also have computed the flavor symmetric channel, for which the interpolating operator is given by,
\begin{align}\label{eq:sixq_1}
\mathcal{H}^{S}_{Qql}&= \frac{1}{\sqrt{3}}\Big( 3 \big[QlqQlq \big] + \big[lqlQqQ \big] + \big[qlqQlQ \big]  \Big).
\end{align}
One can easily notice that, by construction, these states
are the heavy quark generalisation of the singlet and the 27-plet  of the SU(3) flavor symmetry.
By choosing the appropriate quark flavor for $Q,q$  in Eq.(\ref{eq:hexaquark}) one obtains three possible flavor combinations namely $\mathcal{H}_{bcl}, \mathcal{H}_{bsl}$ and $\mathcal{H}_{csl}$.
We also consider the case where $q=l$, i.e, two degenerate light flavors with isospin symmetry.  These dibaryons will be denoted as $\mathcal{H}_b$ and $\mathcal{H}_c$ corresponding to the case of $Q=b$ and $Q=c$ respectively.

The non-interacting two-baryon thresholds for the above six-quark configurations related to these dibaryons involve both light and heavy single baryons, as will be discussed in the next section. Those single baryon correlators are computed using the standard interpolating operators for single baryons.

\section{\label{sec:results}Analysis and Results}
\begin{figure*}[t]
\includegraphics[height=0.5\textwidth, width=0.8\textwidth]{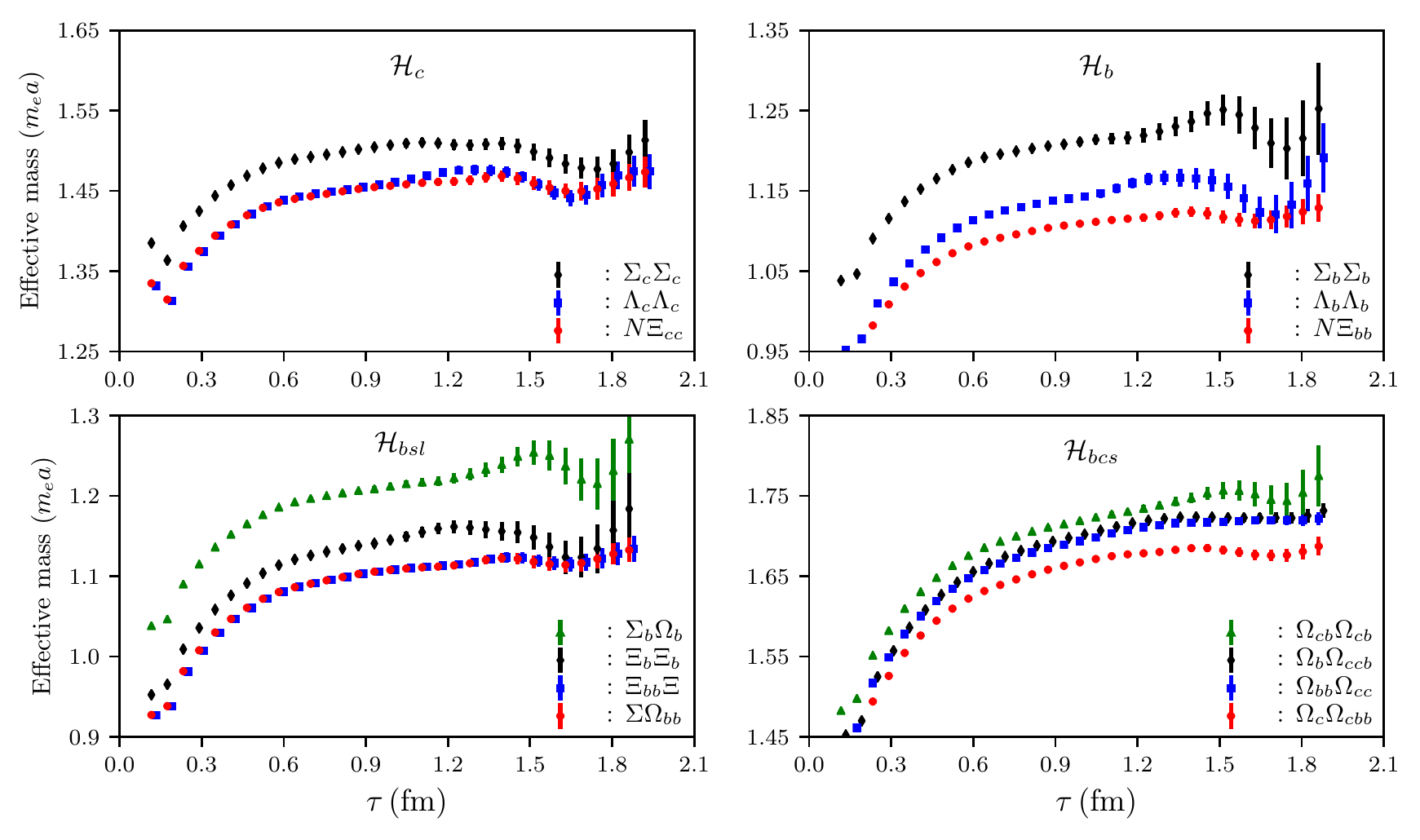}
\caption{\label{fig:thresholds}Effective masses, as defined in Eq.(\ref{eq:eff_mass}), showing the ordering of the non-interacting two-baryon thresholds corresponding to the dibaryons $\mathcal{H}_c$ (top left), $\mathcal{H}_b$(top right), $\mathcal{H}_{bsl}$ (bottom left) and $\mathcal{H}_{bcs}$ (bottom right). The relative position of the lowest non-interacting two-baryon state for each case matches with that shown in Table \ref{tab:threshold}. All effective masses (here and hereafter) are shown in terms of the lattice unit $a = 3390.5$ MeV.}
\end{figure*}

In this section we discuss our calculations and present the results with relevant analysis. With the operators ($\mathcal{O}$) so constructed, using the interpolating fields as given in Eqs.(\ref{eq:hexaquark},\ref{eq:sixq_0},\ref{eq:sixq_1}), 
we compute the single baryon and dibaryon two-point correlation functions between source ($t_i$) and sink ($t_f$) time-slices, 
\begin{align}\label{eq:corr_fun}
C_{\mathcal{O}}(t_i,t_f) = \sum_{\vec{x}} e^{-i\vec{p}.\vec{x}}\langle 0 | \mathcal{O}(\vec{x}_f,t_f)\bar{\mathcal{O}}(\vec{x}_i,t_i)|0 \rangle.
\end{align}
For each case the ground state mass is obtained by fitting the respective average correlation function $ C_{\mathcal{O}}(\tau)$ with a single exponential at sufficiently large times ($\tau = t_f-t_i$). Coulomb gauge fixed wall sources are employed to obtain good overlap to the ground states. The single  baryon correlators are utilized to evaluate the non-interacting two-baryon states.
To evaluate the possible bindings of the dibaryon states it is foremost important to find first the threshold levels and particularly the lowest non-interacting two baryon energy level. Below we discuss that.

\subsection{Threshold energy levels}
We first discuss the relevant thresholds for the charm and bottom dibaryons, $\mathcal{H}_c(cudcud)$  and $\mathcal{H}_b(budbud)$. To represent the correlation functions $C(\tau)$ showing their signal saturation, signal-to-noise ratio and possible fit ranges, we calculate the effective masses as defined below
\begin{align}\label{eq:eff_mass}
m_{eff} = log\Bigl[{{C(\tau)}\over {C(\tau+1)}}\Bigr]. 
\end{align}
Figure~\ref{fig:thresholds} shows the representative effective mass plots of the various possible non-interacting two-baryon  correlators ($C_{T}(t)$). These are obtained from the separate two baryon ($B_1$ and $B_2$) correlators as,
\begin{equation}\label{eq:thr}
C_{T}(\tau) = C_{B_{1}}(\tau) \times  C_{B_{2}}(\tau) .
\end{equation}
The top left plot in Figure~\ref{fig:thresholds}  shows the effective masses of possible non-interacting two-baryon threshold states corresponding to $\mathcal{H}_c$, namely $\Sigma_c \Sigma_c$, $\Lambda_c \Lambda_c$, and $N \Xi_{cc}$, color coded in black (diamond), blue (square) and red (circle), respectively. Similarly, the top right plot
shows the effective masses of $\Sigma_b \Sigma_b$, $\Lambda_b \Lambda_b$ and $N \Xi_{bb}$ for the possible non-interacting two baryon states corresponding to the dibaryon $\mathcal{H}_{b}$.
In both cases, the results are computed at the SU(3) symmetric point, which corresponds to $am_l = am_s = 0.028$ for the $48^3\times 144$ lattice with the lattice spacing $a = 0.0582$ fm.
In both cases, the lowest threshold can be seen to be that of $N\Xi_{QQ}$.
This is in contrast with the $H$ dibaryon case, where the lowest threshold is that of the two non-interacting $\Lambda$ baryons.
In addition, the splitting between the $\Lambda_Q \Lambda_Q$ and the $N \Xi_{QQ}$ increases as the heavy quark mass becomes heavier $-$ from the charm quark to the bottom quark. This is also consistent with the known experimental results for heavy baryons~\cite{Zyla:2020zbs}, and lattice determination of single baryons~\cite{Brown:2014ena, Mathur:2018rwu, PhysRevLett.121.202002} where experimental results are not available. Taken together experimental values of light and charmed baryons and lattice extracted values for bottom baryons, one arrives at the following numbers at the physical quark masses \cite{Zyla:2020zbs, Brown:2014ena, Mathur:2018rwu, PhysRevLett.121.202002}: \begin{eqnarray}
 M(\Lambda\Lambda) - M(N \Xi) &=& - 21.7  \, \hbox{MeV} \nonumber \\
 M(\Lambda_c\Lambda_c) -  M(N \Xi_{cc}) &=& 13.05 \, \hbox{MeV} \nonumber\\
 M(\Lambda_b\Lambda_b) - M(N \Xi_{bb}) &=& 158(35) \, \hbox{MeV}
\end{eqnarray}
That is, whereas the lowest threshold state is $\Lambda\Lambda$ for the light H-dibaryon, the lowest threshold states for $\mathcal{H}_c$ and $\mathcal{H}_b$ are $N \Xi_{cc}$ and $N \Xi_{bb}$, respectively.
We also further point out that in the literature for searches of bound heavy charm dibaryons~\cite{Meguro:2011nr}, the ground state of the dibaryon is often compared incorrectly with the $\Lambda_c \Lambda_c$ threshold instead of the  correct threshold $N \Xi_{cc}$.  The $\Sigma_Q \Sigma_Q$ thresholds in both cases turn out to be higher in energy, similar to the SU(3) case of the $H$ dibaryon.

Similarly, in the bottom two plots of Fig~\ref{fig:thresholds} we show the representative effective masses of the various possible non-interacting thresholds for the dibaryons $\mathcal{H}_{bsl}$ and $\mathcal{H}_{bcs}$. Here the possible elastic thresholds for $\mathcal{H}_{bsl}$ are the two-baryons  $\Xi\, \Xi_{bb}, \Sigma \Omega_{bb},  \Xi_b \Xi_b, \Sigma_b \Omega_{b}$,  and for  $\mathcal{H_{bcs}}$, those are $\Omega_{c}\Omega_{cbb}, \Omega_{cc} \Omega_{bb}, \Omega_{cb}\Omega_{cb}, \Omega_{b}\Omega_{ccb} $. From the ordering of states it is clear that the lowest elastic thresholds for these dibaryons are  $\Xi\, \Xi_{bb}$ and $\Omega_{c}\Omega_{cbb}$, respectively.

We identify the lowest thresholds of other dibaryons using the experimental values of the single baryons~\cite{Zyla:2020zbs} as well as lattice-determined values of them when experimental results are not available, particularly for the bottom baryons \cite{Brown:2014ena, Mathur:2018rwu, PhysRevLett.121.202002}. 
In Table \ref{tab:threshold}, we tabulate all the possible elastic threshold states for the dibaryons that we study in this work. The second column shows the possible non-interacting two-baryon states in ascending order of energy and the third column shows the lowest threshold state. 
\begingroup
\renewcommand*{\arraystretch}{1.5}
\begin{table}[h]
\centering
\begin{tabular}{ccc} \hline \hline
  Dibaryon & Possible & Lowest\\
  &Thresholds & Threshold \\ \hline \hline
$H$ & $\Lambda\Lambda, N \Xi, \Sigma \Sigma$ & $\Lambda\Lambda$ \\ \hline
$\mathcal{H_c}$ & $N \Xi_{cc}, \Lambda_c\Lambda_c, \Sigma_c \Sigma_c$ & $N \Xi_{cc}$ \\\hline
$\mathcal{H_b}$ & $N \Xi_{bb}, \Lambda_b\Lambda_b,\Sigma_b \Sigma_b$ & $N \Xi_{bb}$\\ \hline
$\mathcal{H_{csl}}$ & $\Sigma \Omega_{cc},  \Xi\, \Xi_{cc}, \Xi_c \Xi_c, \Sigma_c \Omega_{c}$ & $\Sigma \Omega_{cc}$\\ \hline  
$\mathcal{H_{bsl}}$  & $\Xi\, \Xi_{bb}, \Sigma \Omega_{bb},  \Xi_b \Xi_b, \Sigma_b \Omega_{b}$ & $\Xi\, \Xi_{bb}$\\ \hline 
$\mathcal{H_{bcl}}$ &$\Sigma_{c}\Omega_{cbb}, \Xi_{cc} \Xi_{bb}, \Sigma_{b}\Omega_{ccb}, \Xi_{cb} \Xi_{cb} $& $\Sigma_{c}\Omega_{cbb}$\\ \hline 
$\mathcal{H_{bcs}}$ &$\Omega_{c}\Omega_{cbb}, \Omega_{cc} \Omega_{bb}, \Omega_{cb}\Omega_{cb}, \Omega_{b}\Omega_{ccb} $& $\Omega_{c}\Omega_{cbb}$\\ \hline \hline
\end{tabular}
\caption{\label{tab:threshold}{Lowest and the possible other non-interacting two-baryon states for $H$ and heavy $\mathcal{H}$-dibaryons.}}
	\end{table} 
\endgroup

The non-interacting two-baryon threshold energy levels ($E_{T}$) corresponding to the two-baryon combinations of Table \ref{tab:threshold}, for a given quark mass combinations ($q_1q_2q_3$), are calculated by adding the single baryon masses extracted at those quark masses,
\begin{equation}
E_{T}(q_1q_2q_3) = M_{B_{1}} (q_1q_2q_3)+ M_{B_{2}}(q_1q_2q_3).
\end{equation}
The extracted single baryon masses are found to be consistent with our previous calculations in Refs.~\cite{PhysRevLett.121.202002, Mathur:2018rwu}.

\begin{figure}[t]
\includegraphics[height=0.3\textwidth, width=0.46\textwidth]{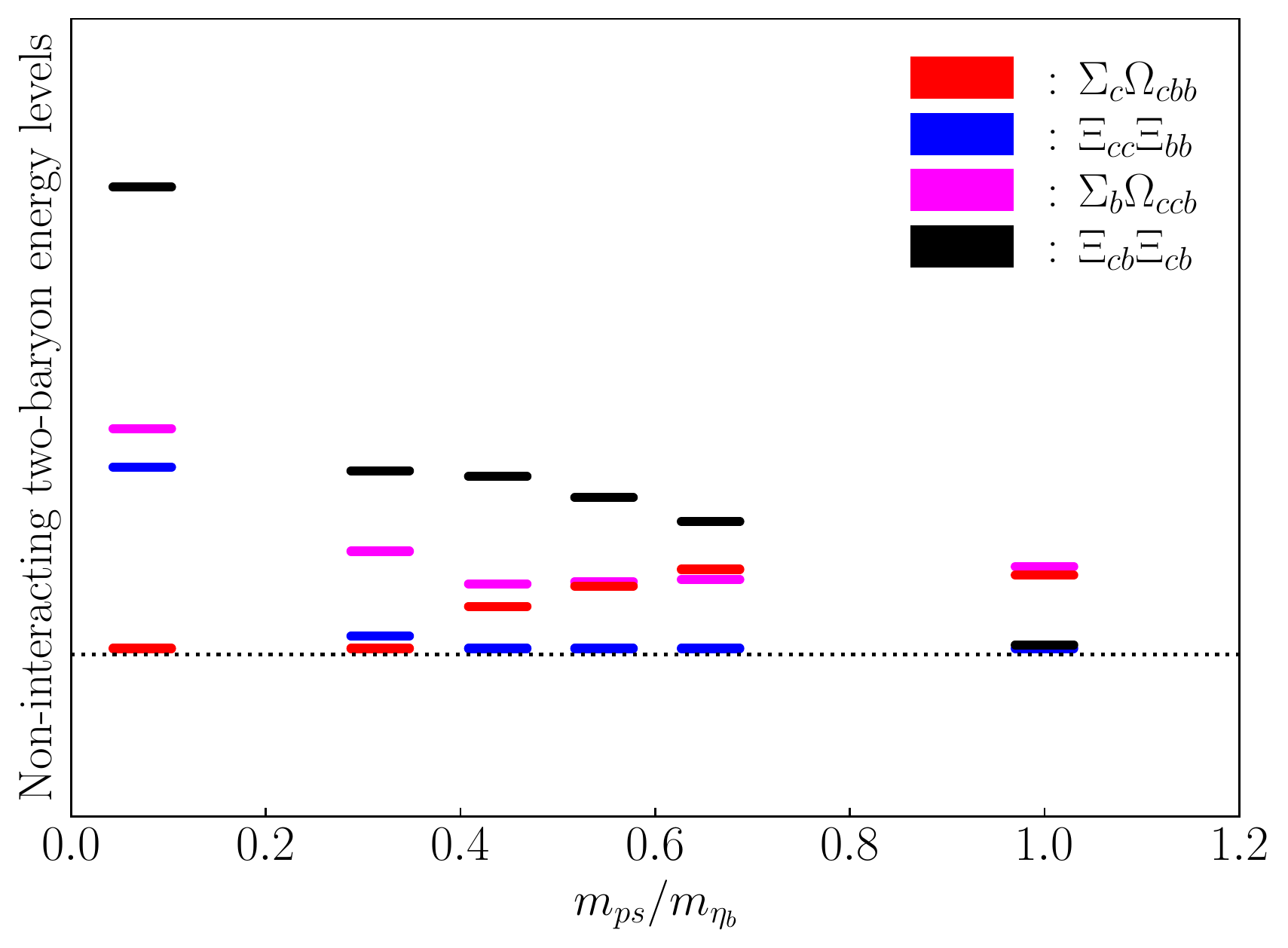}
\caption{\label{fig:th_plot} The relative positions of the energy levels of the non-interacting two-baryons states, corresponding to the dibaryons $\mathcal{H}_{bcl}$, are shown at various quark masses, from light to heavy (shown in terms of ratio of pseudoscalar meson mass ($m_{ps}$) to the $\eta_b$ mass). The lowest thresholds are kept at the same energy level (dashed line) for comparison. The change in the two-baryon particle content of the lowest threshold is clearly visible as one moves from light to heavy quark masses.} 
\end{figure}

As mentioned previously, the lowest two-baryon non-interacting states at light quarks and heavy quarks are different. This is illustrated in Figure \ref{fig:th_plot} for the possible non-interacting two-baryon states for the dibaryon $\mathcal{H}_{bcl}$. We show the variation in terms of the ratio of pseudoscalar meson mass at a quark mass to the $\eta_b$ mass. For comparison purpose we keep the lowest threshold for all cases at the same level but maintain the relative energy differences between various thresholds. While below the charm quark mass the lowest and highest threshold states are $\Sigma_c\Omega_{cbb}$ and $\Xi_{cb}\Xi_{cb}$, respectively, it is completely opposite at the bottom quark mass. Similar level crossings are also found for other dibaryon threshold energy levels. It is therefore crucial to identify  the relative positions of threshold energy levels at a given quark mass for studying heavy dibaryons. It will be very interesting to find a phenomenological explanation behind the minimization of total energy of these threshold levels leading to the observed ordering as shown in the Figure ~\ref{fig:th_plot}

\subsection{Calculation of energy differences}
The ground state energies ($E_{\mathcal{H}}$) of the dibaryons are obtained by fitting the correlators constructed with the operators as mentioned in
Eqs.(\ref{eq:hexaquark},\ref{eq:sixq_0},\ref{eq:sixq_1}), with a single exponential form at large times: $C(\tau) \sim e^{-E_0\tau}$. We then calculate the energy difference ($\Delta E_{\mathcal{H}}$) between the ground state energy of a dibaryon ($E^0_{\mathcal{H}}$) and the elastic threshold energy level ($E^0_T$) as,
\begin{equation}\label{eq:DeltaE}
\Delta E_{\mathcal{H}} = E^0_{\mathcal{H}} - E^0_T .
\end{equation}
For each dibaryon, the whole process of calculating $\Delta E_{\mathcal{H}}$ is performed through a bootstrap method.

We also calculate these energy differences, $\Delta E_{\mathcal{H}}$,  by taking the ratio of the dibaryon correlators  to the two-baryon correlators, as
\begin{equation}\label{eq:ratioE}
R(\tau) = {C_{\mathcal{H}}(\tau) \over  C_{B_{1}}(\tau) \times  C_{B_{2}}(\tau)} = \mathcal{A} e^{-\Delta E_{\mathcal{H}}t} + ...
\end{equation}
A fitting to the above ratio-correlator can also give the energy difference $\Delta E_{\mathcal{H}}$.
While such a ratio-correlator offers the advantage of reducing the
systematic errors, one must be careful in using it as it can possible produce a fake plateau in $R(\tau)$ due to the saturation of different energy states at different time windows. We therefore mostly extract the $\Delta E_{\mathcal{H}}$ values through direct fitting of individual correlators Eq.(\ref{eq:DeltaE}) and the ratio correlators method Eq.(\ref{eq:ratioE}) is used for consistency checks. 

We now present our results for these heavy dibaryons.


\subsection{Flavor-antisymmetric $\mathcal{H}_{c}$ and $\mathcal{H}_{b}$}
\begin{figure*}[t]
  \includegraphics[height=0.5\textwidth, width=0.85\textwidth]{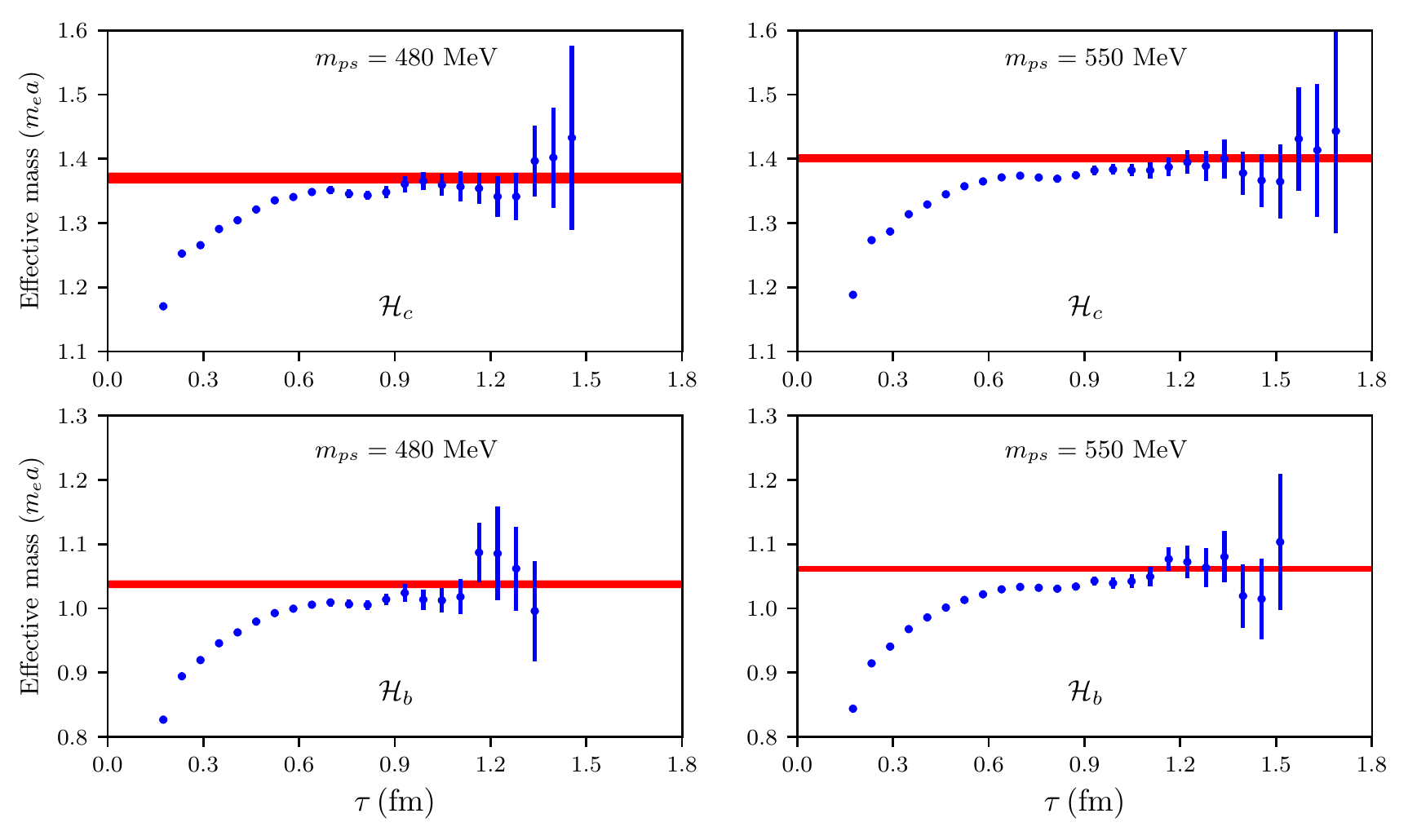}
	\caption{\label{fig:eff_mass_hchb} Effective masses of $\mathcal{H}_{c}(cudcud)$ and $\mathcal{H}_{b}(budbud)$ with the physical charm ($m_c$) and bottom ($m_b$) quark masses and at two values of light quark masses ($m_u = m_d$) corresponding to the pseudoscalar meson masses 480 and 550 MeV. The corresponding two-baryon non-interacting energy levels, extracted at those quark masses, are shown by the horizontal lines. These plots show that the dibaryon and non-interacting two baryon energy levels are consistent to each other signifying the absence of any deeply bound state. These correlation functions are computed at lattice spacing $a = 0.0582$ fermi.}
\end{figure*}

We first present our results for the flavor anti-symmetric $\mathcal{H}_{c}(cudcud)$ and $\mathcal{H}_{b}(budbud)$ dibaryons.
In Figure~\ref{fig:eff_mass_hchb} we show the representative effective masses of $\mathcal{H}_{c}$ (top row) and  $\mathcal{H}_{b}$ (bottom row)  at two light quark masses corresponding to the pseudoscalar masses 480 and 550 MeV. The lowest threshold energy levels extracted from the two baryon non-interacting states, $N\Xi_{cc}$ and $N\Xi_{bb}$, at those quark masses, for $\mathcal{H}_{c}$ and $\mathcal{H}_{b}$ respectively, are shown by the red horizontal lines. It is evident that both the dibaryon energy levels, within the statistical error, overlap with their respective two-baryon non-interacting states at large times.
Correspondingly $\Delta E$, between the lowest energy state  and the two-baryon elastic thresholds $N \Xi_{QQ}$, is found to be consistent with zero for all the light quark masses considered here.
At even lighter quark masses signal-to-noise for the dibaryon correlation functions found to be much poorer and with the large error they overlap further more with the two-baryon non-interacting energy levels. Because of the large statistical error we do not include the data below the 480 MeV pion mass in this pilot study, and hence are unable to conclude on the relative positions and nature of these dibaryon energy levels with respect to their respective thresholds at the physical quark masses.
However, given the trend of the results that we find from higher to the lower quark mass it is highly unlikely that there are any deeper bound state  at the lighter quark masses both for the three-flavored $\mathcal{H}_{c}(cudcud)$ and $\mathcal{H}_{b}(budbud)$ dibaryons. 

\subsection{Flavor-antisymmetric $\mathcal{H}_{csl}, \mathcal{H}_{bsl}, \mathcal{H}_{bcl}$ and $\mathcal{H}_{bcs}$}

\begin{figure*}[h]
  \includegraphics[height=0.5\textwidth, width=0.85\textwidth]{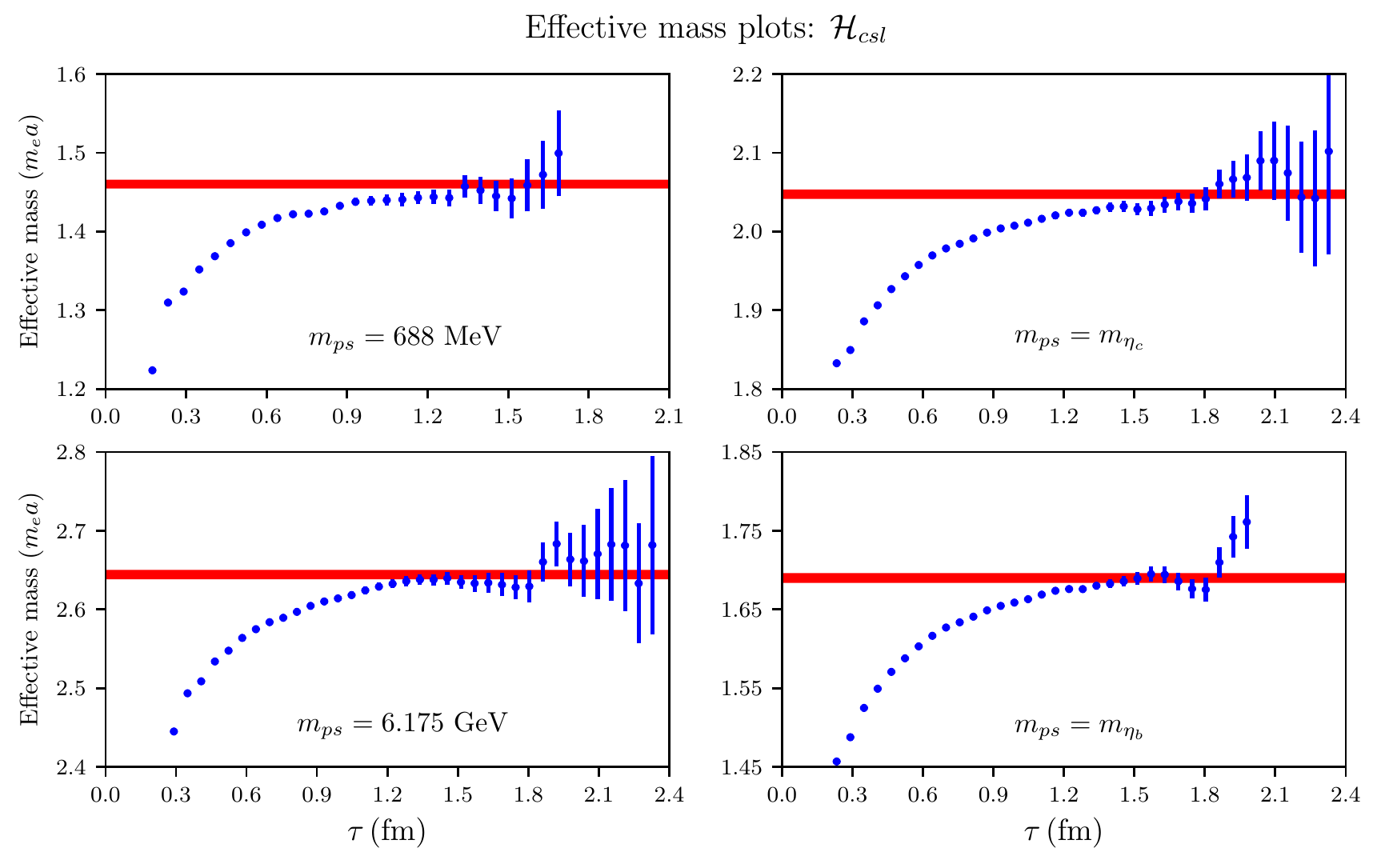}
	\caption{\label{fig:eff_mass_csl} Effective masses of $\mathcal{H}_{csl}$ dibaryons with the physical strange ($m_s$) and charm quark masses ($m_c$) but at various values of $m_l$ corresponding to the pseudoscalar meson masses from 688 MeV to that of $\eta_b$. The various threshold energy levels are shown by the horizontal lines and the ground state of these dibaryons states are consistent to those. Correlation functions are computed at lattice spacing $a = 0.0582$ fermi. }
\end{figure*}

\begin{figure*}[h]
\includegraphics[height=0.5\textwidth, width=0.85\textwidth]{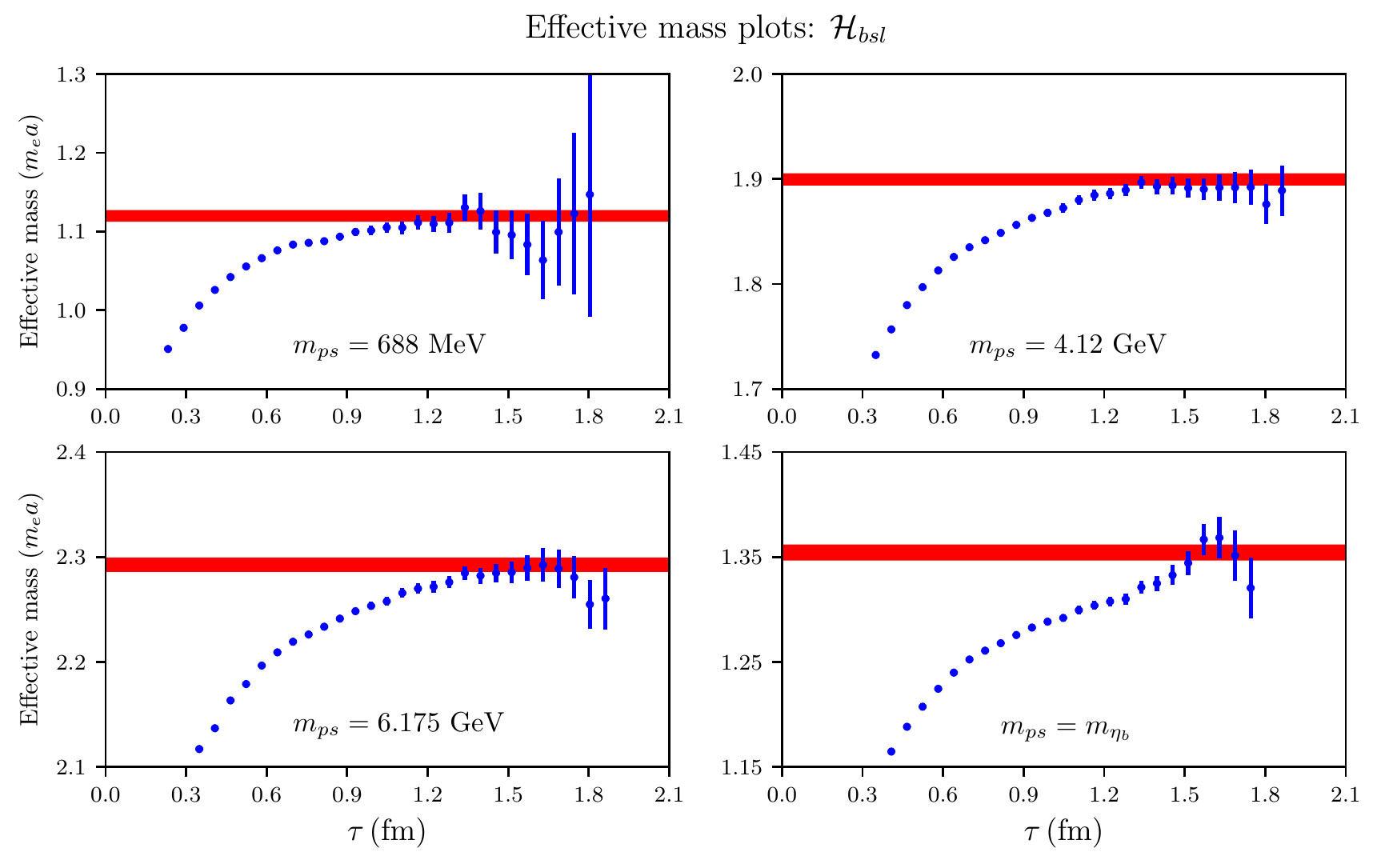}
  \caption{\label{fig:eff_mass_bsl} Effective masses of $\mathcal{H}_{bsl}$ dibaryons with the physical strange ($m_s$) and bottom quark masses ($m_c$) but at various values of $m_l$ corresponding to the pseudoscalar meson masses from 688 MeV to that of $\eta_b$. The various threshold energy are shown by the horizontal lines. The ground state energy of these dibaryons are found to be mostly consistent to these threshold energy levels. Correlation functions are computed at lattice spacing $a = 0.0582$ fermi. }
\end{figure*}

\begin{figure*}[t]
\includegraphics[height=0.5\textwidth, width=0.85\textwidth]{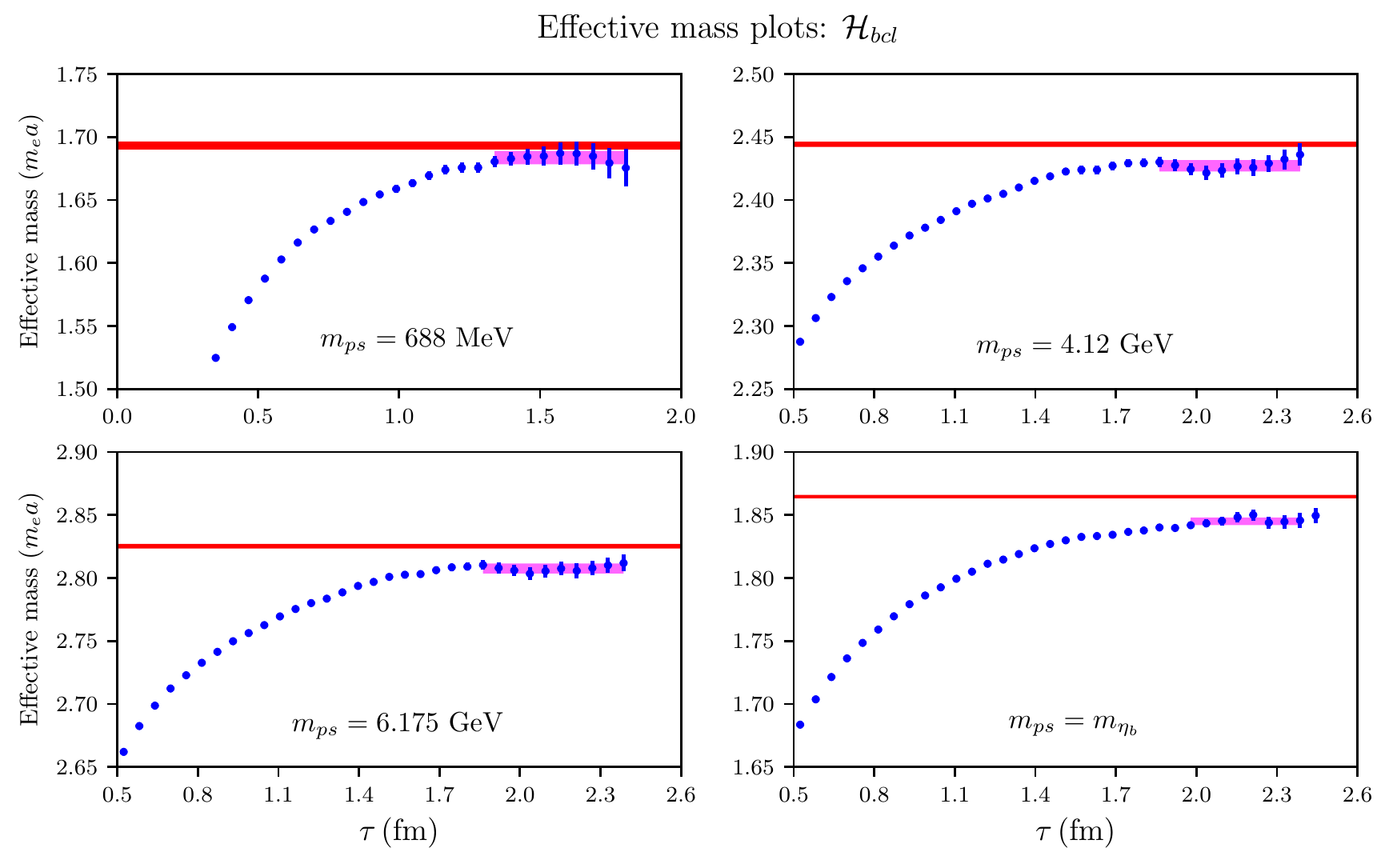}
  \caption{\label{fig:eff_mass_bcl} Effective masses of $\mathcal{H}_{bcl}$ dibaryons with the physical charm ($m_c$) and bottom quark masses ($m_b$) but at various values of $m_l$ ($m_s \le m_l \le m_b$) corresponding to the pseudoscalar meson masses from 688 MeV to that of $\eta_b$. The various threshold energy levels are shown by the horizontal lines. The extracted energy levels (magenta bands) of these dibaryons are found to be mostly below to these threshold energy levels. Correlation functions are computed at lattice spacing $a = 0.0582$ fermi.}
\end{figure*}

In Figure~\ref{fig:eff_mass_csl}, we show the representative effective masses of $\mathcal{H}_{csl}(cslcsl)$; $l \subset u,d$, where the strange and the charm quark masses are set to their physical values whereas the light quark mass ($m_l$) is varied. The pseudoscalar meson masses corresponding to these light quark masses are shown inside the figures.  The non-interacting two-baryon state that is lowest in energy is $\Sigma_{l} \Omega_{cc}$. At each light quark mass ($m_l$) we compute its energy by adding the baryon masses of $\Sigma(lls)$ and $\Omega(ccs)$, and the thresholds thus obtained are represented by the horizontal lines.
We find that for all ranges of $m_l$ the effective masses of $\mathcal{H}_{csl}$, at large times, are either consistent with the two-baryon non-interacting state or stay above that. The effective masses for the cases with $m_l < m_s$ are found to be quite noisy with large errorbars and are not shown here. Within the large statistical error above time-slices 1.5 fm, they overlap with the thresholds indicating the absence of any energy level much below the thresholds.  Here again, it is very likely that the physical states $\mathcal{H}_{csu}$ and $\mathcal{H}_{csd}$ 
are either resonances or a scattering states or loosely bound states near the two-baryon thresholds. To identify that one needs much more statistics and scattering amplitudes analysis of these finite volume energy levels.

In Fig. \ref{fig:eff_mass_bsl}, similarly we show the effective masses of the dibaryons $\mathcal{H}_{bsl}$ where the strange and the bottom quark masses are set to their physical values and the third quark mass ($m_l$) is varied over a range.
Here the elastic threshold state is $\Xi(ssl)\Xi(bbl)$ and is shown by the horizontal line in each plot. At $m_l = m_s$ we find the lowest energy for $\mathcal{H}_{bsl}$ is consistent with the elastic threshold though its central value lies just below that. Here again, at the lighter quark masses we do not find any energy level much below the threshold that we can distinguish from the threshold within the statistics used in this pilot study. It is very likely that there is no deeply bound state of $\mathcal{H}_{bsl}$ dibaryons at the physical quark masses.  However, there are hints of an energy level below the threshold for each unphysically large values of $m_l$ ($ > m_c$), and the energy splitting ($|\Delta E|$) between them increases as $m_l$ increases further.

Next we discuss the dibaryons $\mathcal{H}_{bcl}(bclbcl)$; $l \subset u,d$.
In Fig. \ref{fig:eff_mass_bcl} we show their representative effective masses, where the charm and the bottom quark masses are set to their physical values and the other quark mass ($m_l$) is varied over a range. As shown in Figure \ref{fig:th_plot}, the corresponding lowest thresholds are different at different quark masses and those are shown by the horizontal lines. Here the main observation is that unlike the previous cases, we find an energy level consistently below the threshold, particularly when $m_l$ is large. We fit the  $\mathcal{H}_{bcl}$ correlators with a single exponential and extract its ground state energy $E_{\mathcal{H}_{bcl}}$ at a large time. The fit ranges and fit values with one standard deviation ($\sigma$) are shown by the magenta band. We find it to be consistently lower than the threshold values for
most quark masses for $m_s \le m_l$: $E_{\mathcal{H}_{bcl}} < E^{th}_{bcl}$. The energy difference
$\Delta E = E^{th}_{bcl} - E_{\mathcal{H}_{bcl}}$ increases as the quark mass $m_l$ increases from light to the bottom quark masses.

\begin{figure}[h]
  \hspace*{-0.2in}  
  \includegraphics[height = 0.32\textwidth, width=0.5\textwidth]{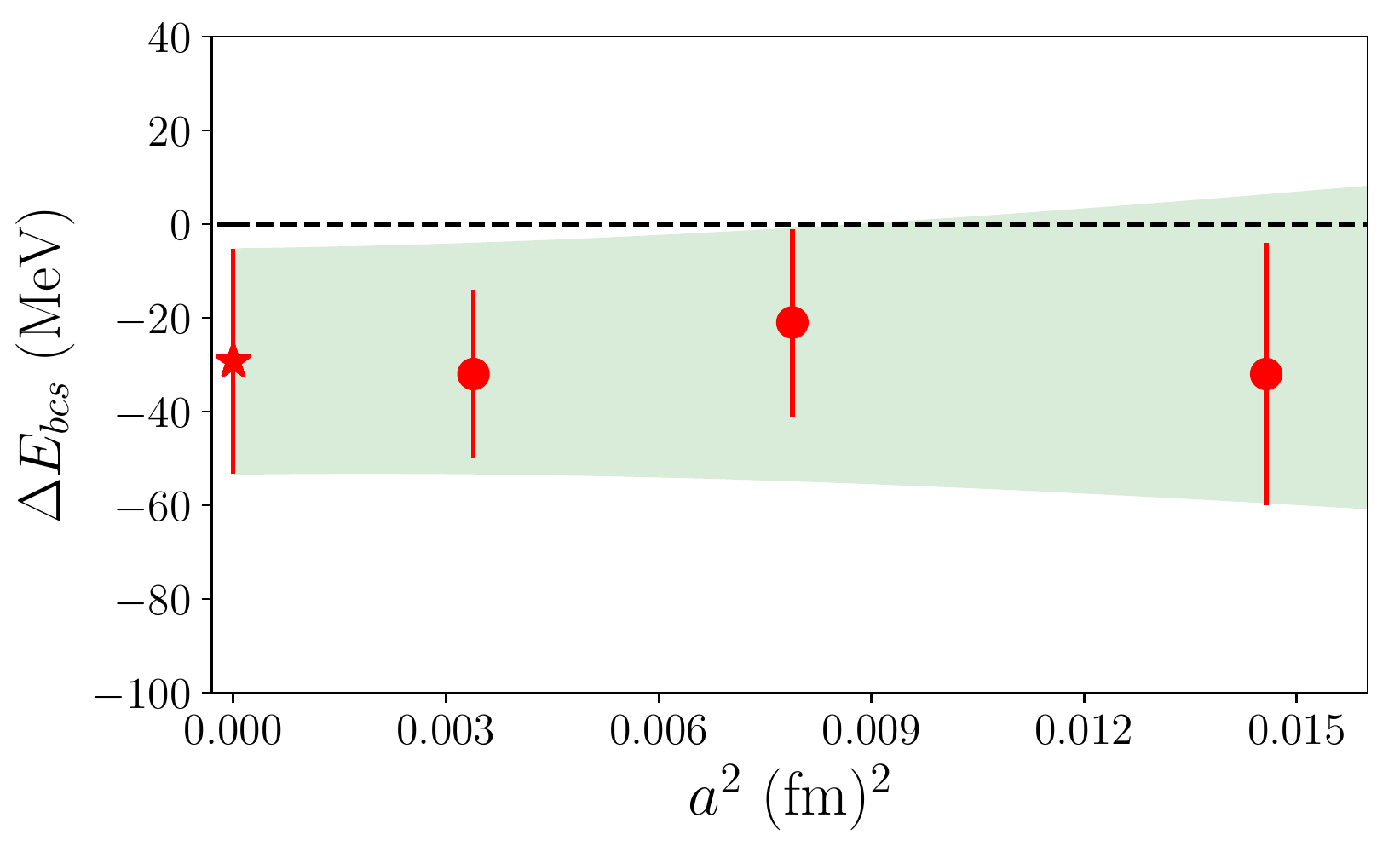}
  \caption{\label{fig:bcs} Continuum extrapolation result for the energy difference ($\Delta E$) between the ground state of the dibaryon $\mathcal{H}_{bcs}$ and the lowest energy level of the corresponding non-interacting two-baryon state  $\Omega_c\Omega_{cbb}$. The point with the star symbol shows the continuum extrapolated value obtained with a form $\Delta E(a) = A + B a^2$, $a$ being the lattice spacing. The green band depicts the fitted one standard deviation error-band.}
\end{figure}

At $m_l = m_s$, the relevant state is the physical three-flavored $\mathcal{H}_{bcs}$ dibaryon which is particularly interesting.  We find that at large times the lowest energy level is below but consistent with the threshold within 1.5-$\sigma$ errorband. We extract the $\Delta E$ value of $\mathcal{H}_{bcs}$ on three lattice spacings and show the results in Fig \ref{fig:bcs}, and also tabulate that in Table \ref{tab:bcs}. A continuum extrapolation with the form $A + Ba^2$  yields a value of $\Delta E_{bcs}|_{\hbox{cont}} = -29(24)$ MeV. This form of extrapolation is justified by the usage of overlap fermions which have no $\mathcal{O}(ma)$ error. One can also use the $\mathcal{O}(a^2log(a))$ term. However, with only three data points, inclusion of such terms is not possible for us.   Though the continuum value is consistent with zero within 1.5-$\sigma$ band, it is clearly noticeable that the ground state energy of  $\mathcal{H}_{bcs}$ is consistently below the non-interacting threshold state $\Omega_{c}\Omega_{bb}$ on all three lattice ensembles. The extrapolated continuum result suggests that the $\mathcal{H}_{bcs}$ dibaryon possibly has non-zero binding. With the given statistics and with fits without considering correlations between the thresholds and dibaryon correlators, it will not be possible to reach a  definite conclusion about the nature of binding for $\mathcal{H}_{bcs}$. Moreover a detail finite volume amplitude analysis of the extracted energy levels is essential to reach a definite conclusion which is beyond the scope of this calculation. Nevertheless, findings from this pilot study are definitely encouraging for pursuing a more quantitative study in the future to achieve that goal.

At the lighter quark masses the signal-to-noise ratio found to be much poorer.
Though the central values of the extracted energy levels for $\mathcal{H}_{bcl}$
dibaryons are always found to be below the lowest threshold, with the given statistics they are consistent with the lowest threshold.  One needs a detail finite volume amplitude analysis with more statistics to find if there is a loosely bound state, or a resonance at threshold or a scattering state for $\mathcal{H}_{bcu}$ and $\mathcal{H}_{bcd}$ dibaryons.
{
\begingroup
\noindent
\renewcommand*{\arraystretch}{1.9}
\setlength{\tabcolsep}{5pt}
\begin{table}[t]
	\centering
	\caption{\label{tab:bcs}{The energy difference ($\Delta E$) between the ground state of the $\mathcal{H}_{bcs}$ dibaryon and the lowest energy level of the non-interacting two-baryon states.  The last column is the continuum extrapolation results from three different lattice QCD ensembles with a form $\Delta E(a) = A + B a^2$.}}
        \vspace*{0.1in}
	\begin{tabular}{c c | c }\hline \hline
	  $a$ & $\Delta E$  & $\Delta E^{bcs}|_{a=0}$ \\
           (fm) & (MeV) & (MeV)\\
	\hline \hline	          
          $0.01207$ & $-32 (28)$ & \\
          $0.0888$  & $ -21 (20)$ & $-29 (24)$ \\
          $0.0582$  & $ -32 (18)$ &  \\
	\hline \hline	
	\end{tabular}
\end{table}
\endgroup 
}

{
\begingroup
\noindent
\renewcommand*{\arraystretch}{1.6}
\setlength{\tabcolsep}{4pt}
\begin{table}[h]
	\centering
	\caption{\label{tab:bcl}{The energy difference ($\Delta E$) between the ground state of the $\mathcal{H}_{bcq}$ dibaryon and the lowest energy level of the non-interacting two-baryon states. The pseudoscalar meson masses ($m_{ps}$) corresponding to various quark masses $m_q$ in between the strange and the bottom quark masses are shown by the first column.}}
	\begin{tabular}{c c }\hline \hline
		$m_{ps}$ &$\Delta E$ \\
		 (MeV) & (MeV)\\ \hline \hline
          $688$ & $-32 (18)$\\
          $2985$ & $-45 (13)$\\
          $5146$ & $-58 (12)$\\
          $6175$ & $-60 (10)$\\
          $9399$ & $-67 (9)$\\
		\hline \hline
	\end{tabular}
\end{table}
\endgroup
}

\begin{figure}[t]
  \hspace*{-0.25in}
	\includegraphics[height = 0.32\textwidth, width=0.5\textwidth]{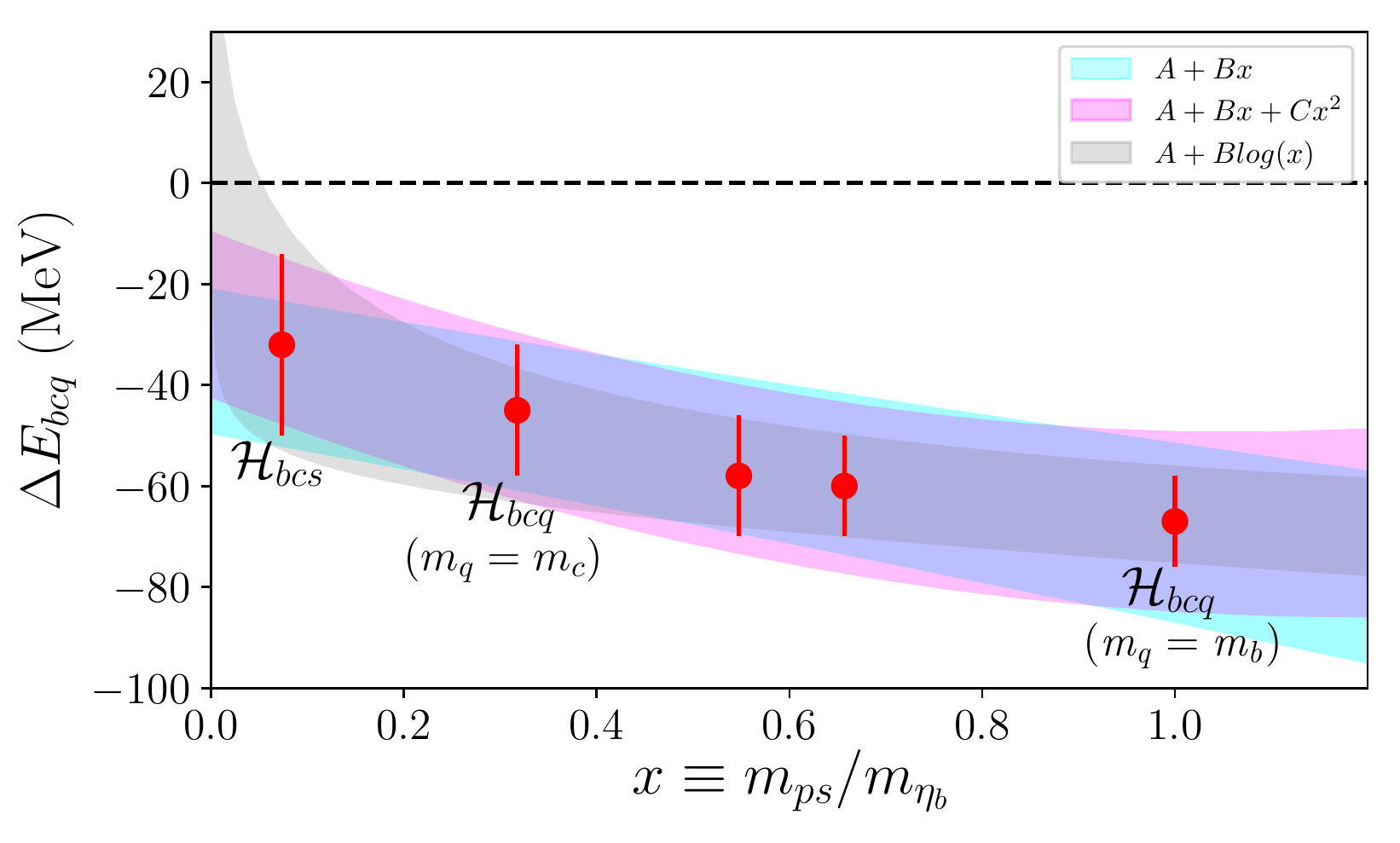}
	\caption{\label{fig:bcq_deltaE}
	The energy difference ($\Delta E$) between the ground state of the $\mathcal{H}_{bcq}$ dibaryons and the lowest energy level of the non-interacting two-baryon states at various values of the quark mass $m_q$ in between the strange to the bottom quarks.  This figure shows that the finite volume $\Delta E$, which is related to the infinite volume binding energy of $\mathcal{H}_{bcq}$, increases with $m_q$, leading to the result that while the physical dibaryon $\mathcal{H}_{bcs}$ is more likely bound or weekly bound, other heavier unphysical dibaryons are more likely strongly bound as $m_q$ increases. Errorbands represent the fitting forms: $\Delta E(x) = A + B x$ (cyan), $\Delta E(x) = A + B x + C x^2$ (magenta) and  $\Delta E(x) = A + B log(x)$ (grey), where $x$ is the ratio of the pseudoscalar meson mass at $m_q$ to the $\eta_b$ mass ($x = m_{ps}/m_{\eta_b}$).}
\end{figure}

\begin{figure}[h]
  \vspace*{-0.1in}
  \includegraphics[height = 0.3\textwidth, width=0.485\textwidth]{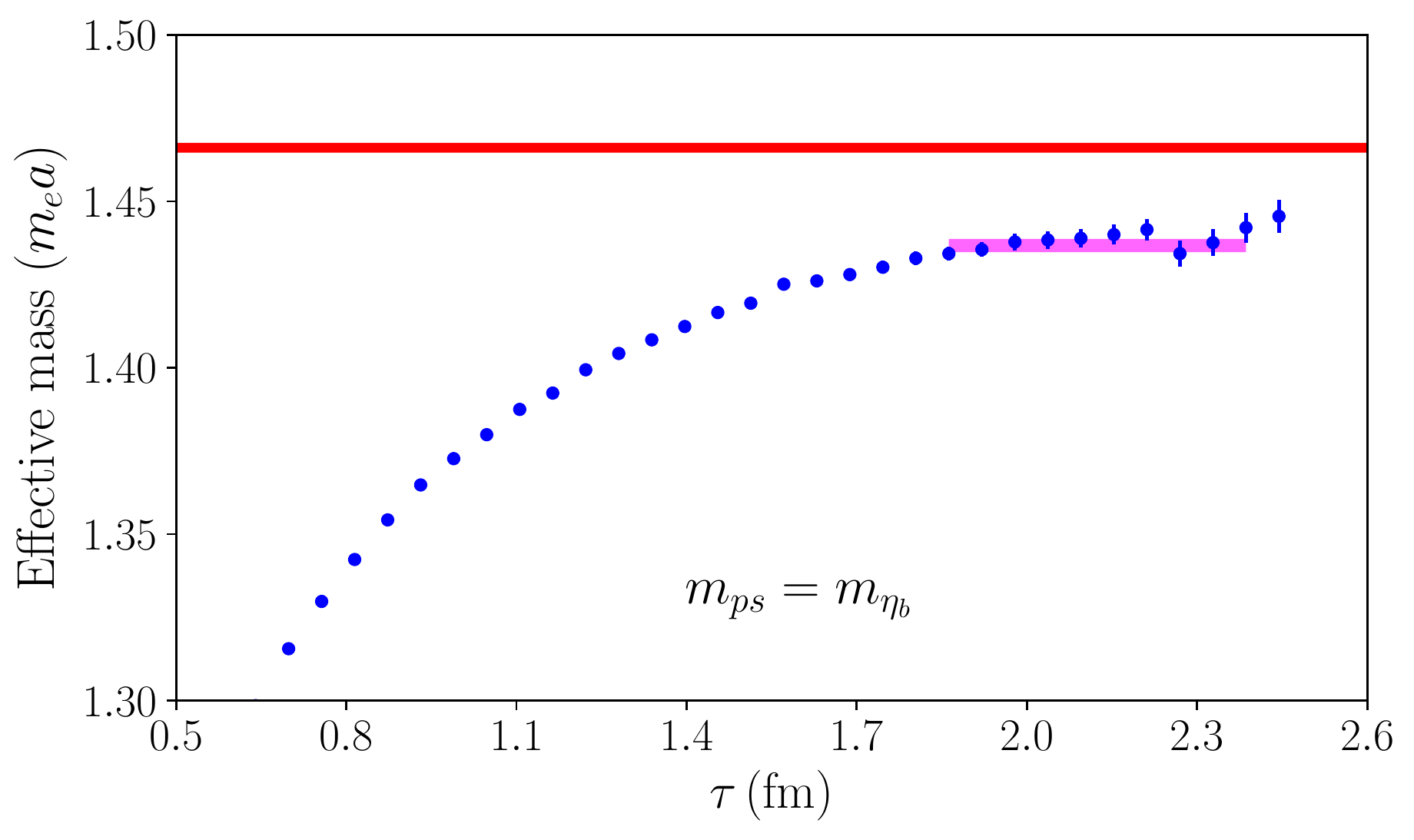}
  \caption{\label{fig:eff_mass_bbb} Effective mass of the very heavy three-flavored dibaryon with all the quark masses set to the bottom quark mass. The energy difference $|\Delta E|$ between the lowest energy level and the elastic threshold is the largest ( $\sim 100$ MeV) in this case indicating the strongest binding of this unphysical three-flavored dibaryon.}
\end{figure}

In Table \ref{tab:bcl} we show $\Delta E$ values for $\mathcal{H}_{bcl}$ at various pseudoscalar meson masses corresponding to the quark masses $m_s \le m_l \le m_b$. In Fig. \ref{fig:bcq_deltaE} we show the variation of $\Delta E_{bcl}$. Following HQET it is expected that the variation of $\Delta E$ will scale with heavy quark masses \cite{Isgur:1991wq}. We thus plot $\Delta E_{bcl}$ as a function of pseudoscalar masses which also scale with quark masses in the heavy quark limit. The $x$-axis is normalized by the mass of $\eta_{b}$ so that at the bottom quark its value becomes 1.   The errorbands shown in the figure are obtained by fitting $\Delta E_{bcl}$ as a function of $x = m_{ps}/\eta_{b}$ with the forms $\Delta E(x) = A + B x$ (cyan), $\Delta E(x) = A + B x+ C x^2$ (magenta) and  $\Delta E(x) = A + B\, log(x)$ (grey). It is interesting to note that a {\it logarithmic} form also fits the data very well which could be phenomenologically interesting to consider for other splittings for their heavy quark mass dependence.

From the above discussion it is quite apparent that when any of the quark mass in a three-flavored dibaryon $\mathcal{H}_{q_1q_2q_3}$ becomes heavier its binding tends to increase. Therefore the strongest binding is expected for the case when $m_{q_1} = m_{q_2} = m_{q_3} = m_b$.  In Fig. \ref{fig:eff_mass_bbb} we show the effective mass plot of this case which shows that the lowest energy state lies much below the corresponding elastic threshold. The extracted $\Delta E$ for this case is found to be $-99(8)$ MeV,
on the fine lattice ensemble. Interestingly, it is consistent with the binding energy, $-109(5)$ MeV, of deuteron-like heavy dibaryons when all quark masses are set at the bottom quark mass \cite{Junnarkar:2019equ}, and also with the binding energy, $-89^{-17}_{+12}(12)$ MeV, of the single flavored heavy dibaryon at the bottom quark mass \cite{Mathur:2022nez}. It indicates that at very heavy quark masses, bindings of the single, two and three-flavored dibaryons are similar.

The conclusions on $\mathcal{H}_{bcl}$ dibaryons are the following: 
\begin{enumerate}[(I)]
\item We find a finite volume energy level below the threshold for $\mathcal{H}_{bcs}$ for all the ensembles used here. A continuum extrapolation yields this energy difference from threshold $\Delta E_{bcs}|_{\hbox{cont}} = -29(24)$ MeV (Table \ref{tab:bcs} and Figure \ref{fig:bcs}). 
\item This energy difference increases as the quark mass $m_l$ increases  and it becomes maximum at the bottom quark mass ($m_l = m_b$) (Table \ref{tab:bcl} and Figure \ref{fig:bcq_deltaE}). 
\item There is no finite volume energy level much below the lowest thresholds
 for the physical $\mathcal{H}_{bcl}; \, l \subset u, d$, dibaryons. However, there is an indication for a finite-volume energy level close to the threshold which we could not resolve and needs to be investigated further to find whether that energy level is associated with a closely bound state, or a resonance at threshold or a scattering state.
\end{enumerate}

\subsection{Flavor-symmetric three-flavored heavy dibaryons}

\begin{figure}[hbt]
  \vspace*{-0.15in}
  \includegraphics[height=0.25\textwidth, width=0.435\textwidth]{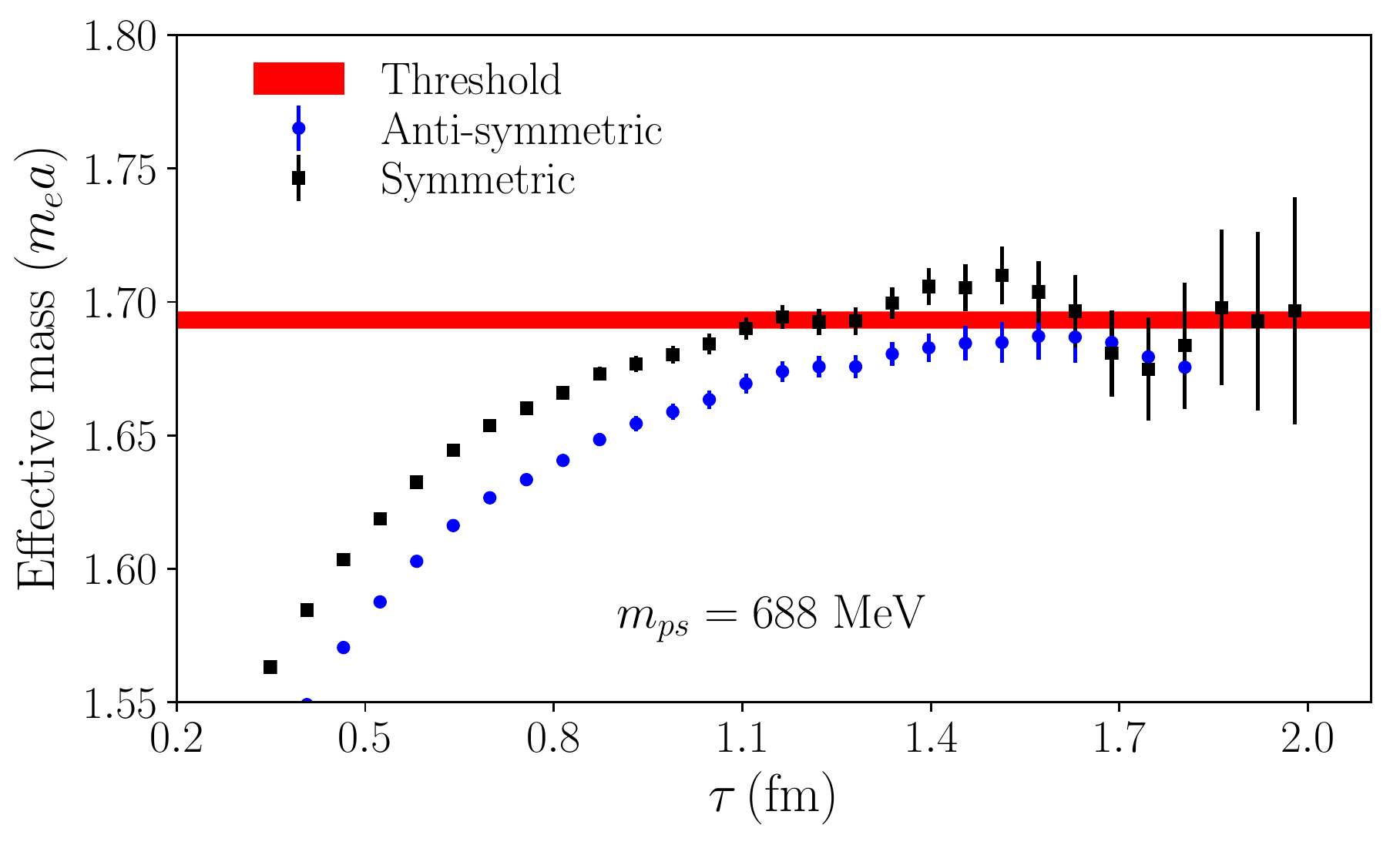}\\
  \includegraphics[height=0.25\textwidth, width=0.45\textwidth]{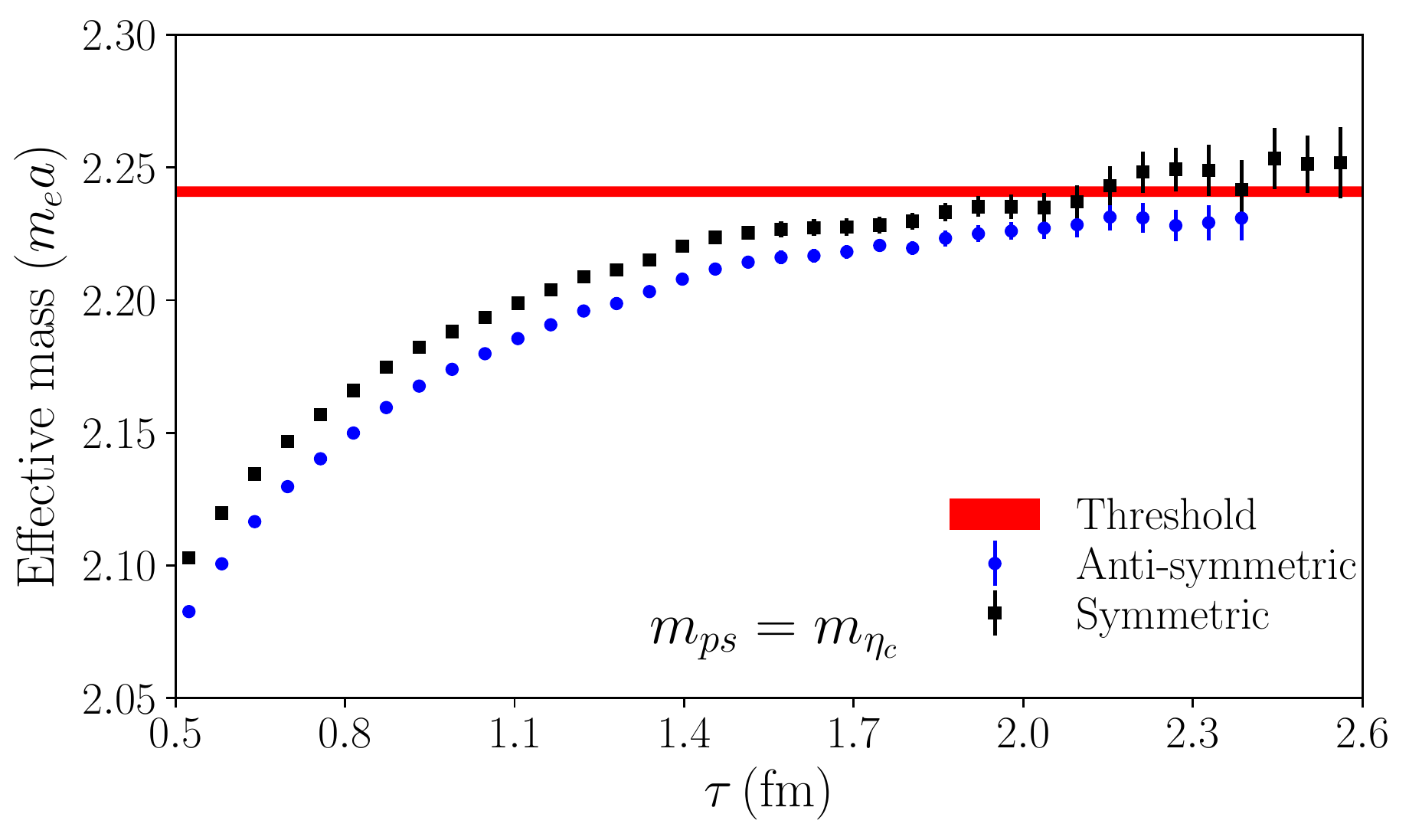}\\
  \includegraphics[height=0.25\textwidth, width=0.45\textwidth]{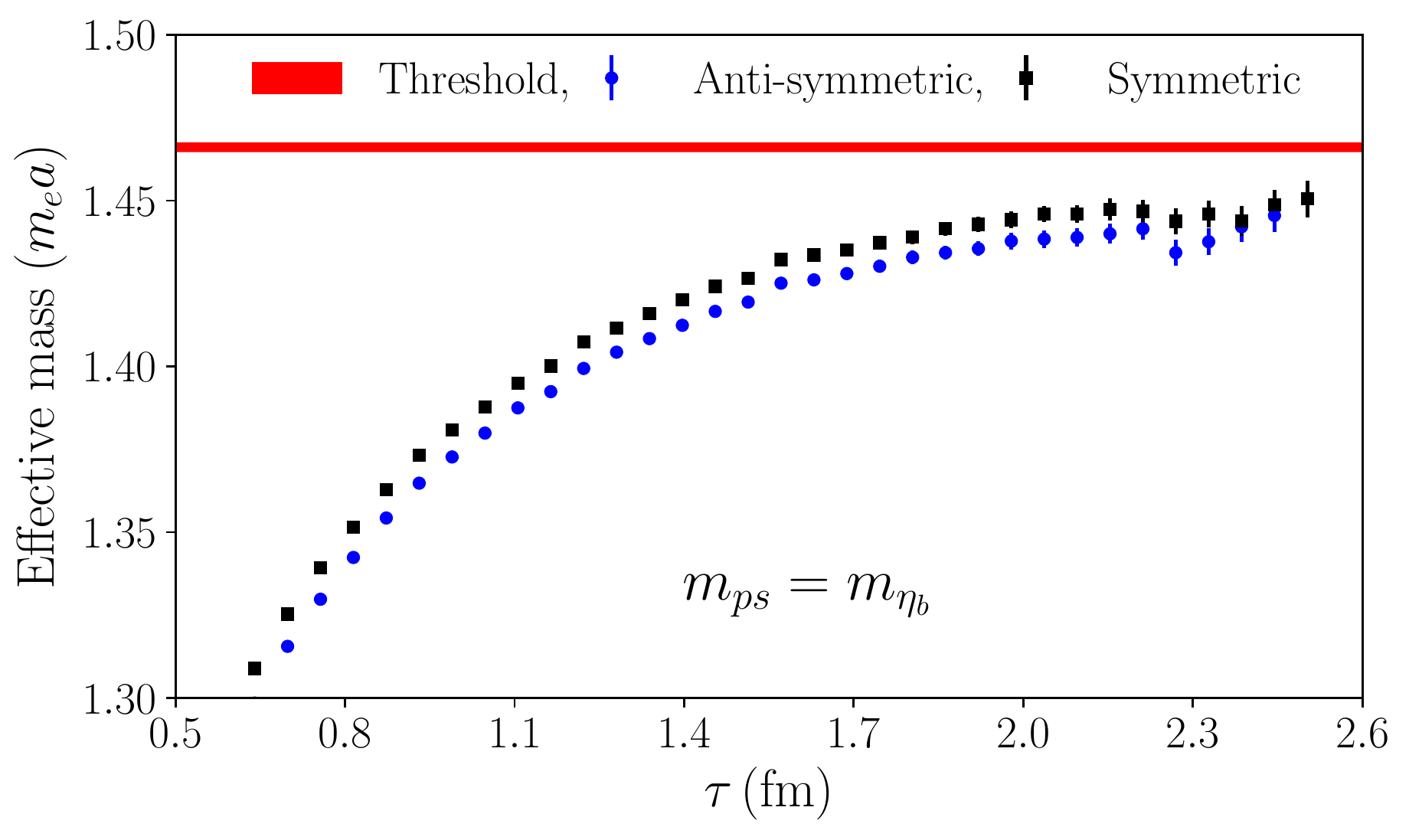}
	\caption{\label{fig:eff_mass_bcl_symm} Effective masses of $\mathcal{H}_{bcl}$ dibaryon for the symmetric flavored cases (black square), in comparison to the anti-symmetric flavored cases (blue circle), at $m_l = m_s$ (top), $m_l = m_c$ (middle), and $m_l = m_b$ on the finest lattice QCD ensemble employed. The threshold energy levels are shown by the horizontal lines. The extracted lowest energy levels for the flavor-symmetric dibaryons are always found to be higher than that of the anti-symmetric cases.}
\end{figure}

We now discuss the results of the flavored-symmetric cases. In Fig. \ref{fig:eff_mass_bcl_symm} we plot the effective masses of the flavored-symmetric $\mathcal{H}_{bcl}$ dibaryon at $m_l = m_s$ (top plot) and $m_l = m_c$ (bottom plot), while keeping the $m_c$ and $m_b$ at their physical values, and compared them with that of the flavored-antisymmetric dibaryons. The data with black squares and blue circles represent the flavored-symmetric and antisymmetric cases, respectively, while the red line represents the non-interacting lowest energy levels of the two baryons. A general feature that we find is that the extracted lowest energy levels for the flavor-symmetric configurations are always found to be higher than that of the antisymmetric cases. We observe that the symmetric $\mathcal{H}_{bcl}$ states, at 
$m_l = m_s$ and $m_l = m_c$, are above their respective elastic threshold energy. Most possibly they are scattering states or resonances above the threshold, and a detailed scattering amplitude analysis of the extracted energy levels is necessary to determine that. However, at the very heavy quark masses, particularly when $m_l = m_c = m_b$, we observe large bindings, but always smaller than the corresponding symmetric cases. For other flavored-symmetric states, for example, for  $\mathcal{H}_{bsl}$ and $\mathcal{H}_{csl}$, we also observe that the lowest energy levels are always higher than their corresponding non-interacting threshold energy levels. Since we do not find any signature of a distinguishable extra energy level below the thresholds for any of the symmetric cases we will not discuss them further in this work.

\subsection{Finite volume effects}
We extracted  dibaryon energy levels on Euclidean lattices at finite volume (3 fermi box extent). These cannot be directly associated with the physical states. In order to do that one needs to perform a finite volume analysis through the scattering amplitude analysis of these finite volume energy levels  \cite{Luscher:1990ck}. However, for multi-hadron states with heavy quarks, it has been noted in Refs~\cite{Junnarkar:2018twb,Junnarkar:2019equ} that the finite volume corrections to infinite volume binding energy of the relevant hadronic state receives a non-trivial large suppression from the masses of the non-interacting heavy hadrons \cite{Beane:2003da,Davoudi:2011md,Briceno:2013bda}, as
\begin{eqnarray}\label{Eq:finite_vol_eff}
 \Delta_{FV} =  E_{FV} - E_{\infty} &\propto &  \mathcal{O}(e^{-k_{\infty} L })/L,\nonumber \\
   \mathrm{with}\quad k_{\infty} &=& \sqrt{(m_1 + m_2) B_{\infty}}\,.
\end{eqnarray}
Here $k_{\infty}$ is the binding momentum of the infinite volume state, $E_{FV}$ is the energy level computed on a cubic lattice, and ($m_1, m_2$) are the masses of the two noninteracting hadrons with the threshold energy $m_1 + m_2$. Therefore, $\Delta_{FV}$ is expected to be smaller for larger values of $m_1$ and $m_2$, that is when the threshold state consists of two heavy hadrons as in the dibaryons that we are studying.
However, as shown in Table \ref{tab:threshold}, for most of these three-flavored dibaryons, one is a light baryon out of the two-baryons at the threshold. In particular, for $\mathcal{H}_c$ and $\mathcal{H}_b$, because of the presence of nucleon at the threshold $(N \Xi_{QQ})$, the combination $m_1 + m_2$ may not provide a stronger suppression in comparison to the case of tetraquarks~\cite{Junnarkar:2018twb}  and two-flavored heavy dibaryons \cite{Junnarkar:2019equ}. Nevertheless, the volume suppression still is expected to be larger than that of light dibaryons. For the cases of $\mathcal{H}_{bcl}$ and $\mathcal{H}_{bcs}$, volume suppression would be even larger. For the unphysical dibaryons as shown in Fig. \ref{fig:bcq_deltaE}, the presence of two heavy baryons will bring back larger volume suppression and one can argue that the energy levels mentioned in Table \ref{tab:bcl} are expected to be closer to their infinite volume limits. Nevertheless, it will be important to perform a finite volume analysis, in particular for $\mathcal{H}_c$ and $\mathcal{H}_b$, as was performed in Ref. \cite{Francis:2018qch}, where infinite volume binding energy was computed by locating the bound state pole in the scattering amplitude \cite{Luscher:1990ck}. Such an analysis is not possible within the framework of our current set up and we would like to pursue that in future.

Beside the statistical and finite volume effects, other systematic errors are also involved in this work, namely, mixed action partially quenching, discretization, scale setting, mass tuning, fit window and electromagnetism. In Ref. \cite{Junnarkar:2019equ} we estimated such errors can be as large as 10 MeV. The parameters set used in this work are similar to that Ref. \cite{Junnarkar:2019equ} and hence we expect similar systematic errors, particularly for the dibaryons $\mathcal{H}_{bcs}$, and those heavier in masses. For other dibaryons involving light quarks these systematics are expected to be larger and without addressing them properly it is not possible to reach a definitive conclusion for their natures of binding.

\section{\label{sec:conclusion} Summary and discussion}
In this work, we report the first lattice QCD study of three-flavored heavy dibaryons both in the flavored-symmetric and antisymmetric channels.
These states are the heavy quark analogues of the much investigated $H$-dibaryon and are in the SU(3) \textbf{27}-plet of quark flavors.
From this pilot study we summarize our findings below.  First, in the flavor-symmetric channel, for the physical dibaryons
$\mathcal{H}_c(cudcud), \mathcal{H}_b(budbud), \mathcal{H}_{csl}(cslcsl), \mathcal{H}_{bsl}(bslbsl)$; $ l \in u,d$, within our statistics we do not find any energy level much below their respective lowest elastic thresholds, which suggests that there is no deeply bound dibaryons in these channels. Most likely they are either loosely bound states near their respective thresholds or resonances just above the thresholds or scattering states. On the other hand, for $\mathcal{H}_{bcs}(bcsbcs)$, we find an energy level below the corresponding lowest non-interacting threshold, $\Omega_c\Omega_{cbb}$. An extrapolation of the energy difference $\Delta E_{\mathcal{H}_{bcs}}$  between the ground state of this dibaryon from  the non-interacting $\Omega_c\Omega_{cbb}$, yields $\Delta E_{\mathcal{H}_{bcs}} = -29 \pm 24$ MeV. Though this result on $\Delta E_{\mathcal{H}_{bcs}}$ has a large error, and is consistent with zero within 1.5 standard deviation, there is a clear trend that it is consistently below the lowest threshold energy level in all the three lattice ensembles employed in this work. Since $\mathcal{H}_{bcs}$ is a physical state and could be an attractive dibaryon candidate to be searched in future at high energy laboratories, this finding of the possibility of an energy level below the threshold from this pilot study is very interesting and calls for an extension with more statistics and better control over systematics.
 However,  for these three-flavored dibaryons, when the light quark mass is set to an unphysically high value, for example for $\mathcal{H}_{bcl}$ with $m_l > m_c$,  while keeping the charm and bottom quark masses at their physical values, we always find an energy level much below the respective threshold energy level. That clearly indicates the possibility of strong binding of those unphysical dibaryons. Moreover, the energy difference from the respective elastic threshold becomes deeper as the quark mass $m_l$ increases, as shown in Fig.~[\ref{fig:bcq_deltaE}]. For the dibaryon $\mathcal{H}_{bsl}$, we also find the presence of an energy level below the elastic threshold at $m_l \sim m_b$, though somewhat closer to the threshold than that of $\mathcal{H}_{bcl}$.  For the dibaryon $\mathcal{H}_{csl}$, within the statistics employed in this study we do not find an energy level below its lowest threshold that can be distinguished from the lowest threshold for any value of quark masses $m_l$ employed in this work. For the flavor-symmetric channels the corresponding energy levels are observed to be always higher than those of flavor-symmetric cases, suggesting possible scattering states or resonances above the thresholds. Taken together all results, we can summarize that for the three-flavored dibaryons $\mathcal{H}_{q_1q_2q_3}$, there is no deeply bound state if any of the quark mass ($m_{q_{i}}$) is below the charm quark mass. However, we find strong indications of a shallow level below the threshold for the physical $\mathcal{H}_{bcs}$ state which needs to be probed further. Moreover, an energy level below the threshold always emerges when all the three quark masses  become heavier than the charm quark mass, and the binding increases with the increase of quark masses.

We would also like to point out that there are different dynamics as far as binding is concerned for the three-flavored light and heavy dibaryons. That is reflected through the presence of different types of two baryons at their respective elastic thresholds. For the $H$-dibaryon, which is the lightest three-flavored dibaryon, the elastic threshold state is $\Lambda\Lambda$ (with $M(\Lambda\Lambda) - M(N \Xi) =  - 21.7$ MeV). This also continues to be the case at the SU(3) point. However, for the heavy three-flavored dibaryons, $\mathcal{H}_{c}$ and $\mathcal{H}_{b}$, the lowest thresholds are $N\,\Xi_{cc}$ and $N\,\Xi_{bb}$, respectively (with $M(\Lambda_c\Lambda_c) - M(N \Xi_{cc}) =  13.05$ MeV and $M(\Lambda_b\Lambda_b) - M(N \Xi_{bb}) = 158(30)$) \cite{Zyla:2020zbs, PhysRevD.90.094507, Mathur:2018rwu, PhysRevLett.121.202002}. The presence of a doubly heavy baryon lowers the threshold for a heavy three-flavored dibaryon.

The results obtained in this work, when are taken together with the findings in doubly-heavy two-flavored deuteron-like dibaryons \cite{Junnarkar:2019equ}, all-heavy single-flavored dibaryons \cite{Mathur:2022nez} and doubly heavy tetraquarks \cite{Bicudo:2012qt,Francis:2016hui,Junnarkar:2018twb,Leskovec:2019ioa}, point to an interesting dynamics of the heavy multi-hadron systems. A common pattern emerges that for a doubly heavy multiquark hadron, the heavier the two heavy quarks the stronger is the binding. However, the mass of other quarks (or antiquarks) towards the strong binding of these systems are quite intriguing.
In the case of doubly heavy tetraquarks, the heavier the heavy quarks (or anti-quarks) and lighter the light antiquarks (or quarks), the stronger is the binding \cite{Francis:2018qch, Junnarkar:2018twb}. On the contrary, for the heavy dibaryons, binding increases when all the quarks are heavier, that is, in the presence of a light quark the binding decreases.
In addition, while for various two-flavored dibaryons with two heavy quarks, the third quark can still be lighter to have an energy level below the elastic threshold,  for the three-flavored case only $\mathcal{H}_{bcs}$ shows such behaviour. All other physical three-flavored $\mathcal{H}$-dibaryons are most likely either unbound or very weakly bound. That is, the two-flavored heavy dibaryons have stronger binding than that of three-flavored heavy dibaryons. However, when all the quarks become much heavier ($m_{q_1,q_2,q_3} \sim m_b$) the one-, two- and three-flavored dibaryons all exhibit similar strong binding. 

The study pursued here is the first effort to investigate the three-flavored heavy dibaryons.  Given the amount of theoretical and experimental efforts put into the exploration of the $H$ dibaryon state, our motivation has been to elucidate the trend of the lattice ground state energy levels with respect to the elastic thresholds as the strange quark becomes heavier.
In doing so, our hope has been to identify a possible favorable three-flavored channels in charm and/or bottom sectors which may exhibit a bound state, and guide in discovering them in future given the large experimental data being collected and to be collected for heavy hadron spectroscopy at various laboratories. This pilot study indicates that the dibaryon $\mathcal{H}_{b}(bcsbcs)$ is possibly such a bound state. Considering the feasibility of discovering it in high energy experimental laboratories, it will thus be worthwhile to pursue a more detailed study in future to get a definite conclusion on the binding of this state. That can be accomplished with the variational method combined with the use of distillation method for dibaryon systems as in Ref. \cite{Francis:2018qch} or with the potential method \cite{Aoki:2012tk}. Along with that, as mentioned earlier, a detailed finite-volume analysis is needed to discern the pole distribution in the scattering amplitude across the complex energy plane. We will pursue such a study in future.

\begin{acknowledgments}
  This work is supported by the Department of Atomic Energy, Government of India, under Project Identification Number RTI 4002. PJ would like to acknowledge support by the Deutsche
Forschungsgemeinschaft (DFG, German Research Foundation) through the CRC-TR 211 ``Strong-interaction matter under extreme conditions''- project number
315477589 - TRR 211. PJ also thanks the GSI Helmholtzzentrum and the TU Darmstadt and its Institut fur Kernphysik for supporting this research. We would also like to thank Shashin Pavaskar and M. Padmanath for useful discussions.
We are thankful to the MILC collaboration and in particular to S. Gottlieb for providing us with the HISQ lattices. Computations are carried out on the Cray-XC30 of ILGTI, TIFR,   and on the Pride/Flock clusters of the Department of Theoretical Physics,
TIFR.   NM would also like to thank Ajay Salve, Kapil Ghadiali and P. M. Kulkarni for computational supports. 
\end{acknowledgments}

\bibliographystyle{utphys-noitalics}
\bibliography{heavy_H}

\providecommand{\href}[2]{#2}\begingroup\raggedright\begin{thebibliography}{10}

\bibitem{Zyla:2020zbs}
{\bfseries Particle Data Group} Collaboration, P.~Zyla {\em et~al.}, ``{Review
  of Particle Physics},'' \href{http://dx.doi.org/10.1093/ptep/ptaa104}{PTEP
  {\bfseries 2020} no.~8, (2020) 083C01}.

\bibitem{Esposito:2014rxa}
A.~Esposito, A.~L. Guerrieri, F.~Piccinini, A.~Pilloni, and A.~D. Polosa,
  ``{Four-Quark Hadrons: an Updated Review},''
  \href{http://dx.doi.org/10.1142/S0217751X15300021}{Int. J. Mod. Phys.
  {\bfseries A30} (2015) 1530002},
\href{http://arxiv.org/abs/1411.5997}{{\ttfamily arXiv:1411.5997 [hep-ph]}}.

\bibitem{Ali:2017jda}
A.~Ali, J.~S. Lange, and S.~Stone, ``{Exotics: Heavy Pentaquarks and
  Tetraquarks},''
\href{http://arxiv.org/abs/1706.00610}{{\ttfamily arXiv:1706.00610 [hep-ph]}}.

\bibitem{Olsen:2017bmm}
S.~L. Olsen, T.~Skwarnicki, and D.~Zieminska, ``{Nonstandard heavy mesons and
  baryons: Experimental evidence},''
  \href{http://dx.doi.org/10.1103/RevModPhys.90.015003}{Rev. Mod. Phys.
  {\bfseries 90} no.~1, (2018) 015003},
  \href{http://arxiv.org/abs/1708.04012}{{\ttfamily arXiv:1708.04012
  [hep-ph]}}.

\bibitem{Guo:2017jvc}
F.-K. Guo, C.~Hanhart, U.-G. Mei\ss{}ner, Q.~Wang, Q.~Zhao, and B.-S. Zou,
  ``{Hadronic molecules},''
  \href{http://dx.doi.org/10.1103/RevModPhys.90.015004}{Rev. Mod. Phys.
  {\bfseries 90} no.~1, (2018) 015004},
  \href{http://arxiv.org/abs/1705.00141}{{\ttfamily arXiv:1705.00141
  [hep-ph]}}.

\bibitem{Karliner:2017qhf}
M.~Karliner, J.~L. Rosner, and T.~Skwarnicki, ``{Multiquark States},''
  \href{http://dx.doi.org/10.1146/annurev-nucl-101917-020902}{Ann. Rev. Nucl.
  Part. Sci. {\bfseries 68} (2018) 17--44},
  \href{http://arxiv.org/abs/1711.10626}{{\ttfamily arXiv:1711.10626
  [hep-ph]}}.

\bibitem{Brambilla:2019esw}
N.~Brambilla, S.~Eidelman, C.~Hanhart, A.~Nefediev, C.-P. Shen, C.~E. Thomas,
  A.~Vairo, and C.-Z. Yuan, ``{The $XYZ$ states: experimental and theoretical
  status and perspectives},''
  \href{http://dx.doi.org/10.1016/j.physrep.2020.05.001}{Phys. Rept. {\bfseries
  873} (2020) 1--154}, \href{http://arxiv.org/abs/1907.07583}{{\ttfamily
  arXiv:1907.07583 [hep-ex]}}.

\bibitem{Liu:2019zoy}
Y.-R. Liu, H.-X. Chen, W.~Chen, X.~Liu, and S.-L. Zhu, ``{Pentaquark and
  Tetraquark states},''
  \href{http://dx.doi.org/10.1016/j.ppnp.2019.04.003}{Prog. Part. Nucl. Phys.
  {\bfseries 107} (2019) 237--320},
  \href{http://arxiv.org/abs/1903.11976}{{\ttfamily arXiv:1903.11976
  [hep-ph]}}.

\bibitem{Belle:2003nnu}
{\bfseries Belle} Collaboration, S.~K. Choi {\em et~al.}, ``{Observation of a
  narrow charmonium-like state in exclusive $B^\pm \to K^\pm \pi^+ \pi^-
  J/\psi$ decays},''
  \href{http://dx.doi.org/10.1103/PhysRevLett.91.262001}{Phys. Rev. Lett.
  {\bfseries 91} (2003) 262001},
  \href{http://arxiv.org/abs/hep-ex/0309032}{{\ttfamily arXiv:hep-ex/0309032}}.

\bibitem{Belle:2011aa}
{\bfseries Belle} Collaboration, A.~Bondar {\em et~al.}, ``{Observation of two
  charged bottomonium-like resonances in Y(5S) decays},''
  \href{http://dx.doi.org/10.1103/PhysRevLett.108.122001}{Phys. Rev. Lett.
  {\bfseries 108} (2012) 122001},
\href{http://arxiv.org/abs/1110.2251}{{\ttfamily arXiv:1110.2251 [hep-ex]}}.

\bibitem{Aaij:2013zoa}
{\bfseries LHCb} Collaboration, R.~Aaij {\em et~al.}, ``{Determination of the
  X(3872) meson quantum numbers},''
  \href{http://dx.doi.org/10.1103/PhysRevLett.110.222001}{Phys. Rev. Lett.
  {\bfseries 110} (2013) 222001},
\href{http://arxiv.org/abs/1302.6269}{{\ttfamily arXiv:1302.6269 [hep-ex]}}.

\bibitem{Aaij:2014jqa}
{\bfseries LHCb} Collaboration, R.~Aaij {\em et~al.}, ``{Observation of the
  resonant character of the $Z(4430)^-$ state},''
  \href{http://dx.doi.org/10.1103/PhysRevLett.112.222002}{Phys. Rev. Lett.
  {\bfseries 112} no.~22, (2014) 222002},
\href{http://arxiv.org/abs/1404.1903}{{\ttfamily arXiv:1404.1903 [hep-ex]}}.

\bibitem{BESIII:2020qkh}
{\bfseries BESIII} Collaboration, M.~Ablikim {\em et~al.}, ``{Observation of a
  Near-Threshold Structure in the $K^+$ Recoil-Mass Spectra in $e^+e^-
  \rightarrow K^+(D_s^-D^{*0}+D_s^{*-}D^0$)},''
  \href{http://dx.doi.org/10.1103/PhysRevLett.126.102001}{Phys. Rev. Lett.
  {\bfseries 126} no.~10, (2021) 102001},
  \href{http://arxiv.org/abs/2011.07855}{{\ttfamily arXiv:2011.07855
  [hep-ex]}}.

\bibitem{LHCb:2021vvq}
{\bfseries LHCb} Collaboration, R.~Aaij {\em et~al.}, ``{Observation of an
  exotic narrow doubly charmed tetraquark},''
  \href{http://arxiv.org/abs/2109.01038}{{\ttfamily arXiv:2109.01038
  [hep-ex]}}.

\bibitem{Aaij:2015tga}
{\bfseries LHCb} Collaboration, R.~Aaij {\em et~al.}, ``{Observation of $J/\psi
  p$ Resonances Consistent with Pentaquark States in $\Lambda_b^0 \to J/\psi
  K^- p$ Decays},''
  \href{http://dx.doi.org/10.1103/PhysRevLett.115.072001}{Phys. Rev. Lett.
  {\bfseries 115} (2015) 072001},
\href{http://arxiv.org/abs/1507.03414}{{\ttfamily arXiv:1507.03414 [hep-ex]}}.

\bibitem{LHCb:2019kea}
{\bfseries LHCb} Collaboration, R.~Aaij {\em et~al.}, ``{Observation of a
  narrow pentaquark state, $P_c(4312)^+$, and of two-peak structure of the
  $P_c(4450)^+$},''
  \href{http://dx.doi.org/10.1103/PhysRevLett.122.222001}{Phys. Rev. Lett.
  {\bfseries 122} no.~22, (2019) 222001},
  \href{http://arxiv.org/abs/1904.03947}{{\ttfamily arXiv:1904.03947
  [hep-ex]}}.

\bibitem{PhysRevLett.112.202301}
{\bfseries WASA-at-COSY Collaboration and SAID Data Analysis Center}
  Collaboration, A.~P. et~al., ``Evidence for a new resonance from polarized
  neutron-proton scattering,''
  \href{http://dx.doi.org/10.1103/PhysRevLett.112.202301}{Phys. Rev. Lett.
  {\bfseries 112} (May, 2014) 202301}.
  \url{https://link.aps.org/doi/10.1103/PhysRevLett.112.202301}.

\bibitem{Molina:2021bwp}
R.~Molina, N.~Ikeno, and E.~Oset, ``{Sequential single pion production
  explaning the dibaryon ''$d^*(2380)$'' peak},''
  \href{http://arxiv.org/abs/2102.05575}{{\ttfamily arXiv:2102.05575
  [nucl-th]}}.

\bibitem{KAPLAN1998329}
D.~B. Kaplan, M.~J. Savage, and M.~B. Wise, ``Two-nucleon systems from
  effective field theory,''
  \href{http://dx.doi.org/https://doi.org/10.1016/S0550-3213(98)00440-4}{Nuclear
  Physics B {\bfseries 534} no.~1, (1998) 329--355}.
  \url{https://www.sciencedirect.com/science/article/pii/S0550321398004404}.

\bibitem{RevModPhys.81.1773}
E.~Epelbaum, H.-W. Hammer, and U.-G. Mei\ss{}ner, ``Modern theory of nuclear
  forces,'' \href{http://dx.doi.org/10.1103/RevModPhys.81.1773}{Rev. Mod. Phys.
  {\bfseries 81} (Dec, 2009) 1773--1825}.
  \url{https://link.aps.org/doi/10.1103/RevModPhys.81.1773}.

\bibitem{Jaffe:1976yi}
R.~L. Jaffe, ``{Perhaps a Stable Dihyperon},''
  \href{http://dx.doi.org/10.1103/PhysRevLett.38.617,
  10.1103/PhysRevLett.38.195}{Phys. Rev. Lett. {\bfseries 38} (1977) 195--198}.
[Erratum: Phys. Rev. Lett.38,617(1977)].

\bibitem{Donoghue:1986zd}
E.~Golowich, J.~Donoghue, and B.~R. Holstein, ``{Weak Decays of the $H$
  Dibaryon},'' \href{http://dx.doi.org/10.1103/PhysRevD.34.3434,
  10.1063/1.36196}{Phys. Rev. {\bfseries D34} (1986) 3434}.
[AIP Conf. Proc.150,952(1986)].

\bibitem{Golowich:1992zw}
E.~Golowich and T.~Sotirelis, ``{O(alpha-s**2) mass contributions to the H
  dibaryon in a truncated bag model},''
\href{http://dx.doi.org/10.1103/PhysRevD.46.354}{Phys. Rev. {\bfseries D46}
  (1992) 354--363}.

\bibitem{Haidenbauer:2011ah}
J.~Haidenbauer and U.-G. Meissner, ``{To bind or not to bind: The H-dibaryon in
  light of chiral effective field theory},''
  \href{http://dx.doi.org/10.1016/j.physletb.2011.10.070}{Phys. Lett. B
  {\bfseries 706} (2011) 100--105},
  \href{http://arxiv.org/abs/1109.3590}{{\ttfamily arXiv:1109.3590 [hep-ph]}}.

\bibitem{Haidenbauer:2015zqb}
J.~Haidenbauer, U.-G. Mei\ss{}ner, and S.~Petschauer, ``{Strangeness S =
  \ensuremath{-}2 baryon\textendash{}baryon interaction at next-to-leading
  order in chiral effective field theory},''
  \href{http://dx.doi.org/10.1016/j.nuclphysa.2016.01.006}{Nucl. Phys. A
  {\bfseries 954} (2016) 273--293},
  \href{http://arxiv.org/abs/1511.05859}{{\ttfamily arXiv:1511.05859
  [nucl-th]}}.

\bibitem{Takahashi:2001nm}
H.~Takahashi {\em et~al.}, ``{Observation of a (Lambda Lambda)He-6 double
  hypernucleus},''
\href{http://dx.doi.org/10.1103/PhysRevLett.87.212502}{Phys. Rev. Lett.
  {\bfseries 87} (2001) 212502}.

\bibitem{Kim:2013vym}
{\bfseries Belle} Collaboration, B.~H. Kim {\em et~al.}, ``{Search for an
  $H$-dibaryon with mass near $2m_\Lambda$ in $\Upsilon(1S)$ and $\Upsilon(2S)$
  decays},'' \href{http://dx.doi.org/10.1103/PhysRevLett.110.222002}{Phys. Rev.
  Lett. {\bfseries 110} no.~22, (2013) 222002},
\href{http://arxiv.org/abs/1302.4028}{{\ttfamily arXiv:1302.4028 [hep-ex]}}.

\bibitem{NPLQCD:2010ocs}
{\bfseries NPLQCD} Collaboration, S.~R. Beane {\em et~al.}, ``{Evidence for a
  Bound H-dibaryon from Lattice QCD},''
  \href{http://dx.doi.org/10.1103/PhysRevLett.106.162001}{Phys. Rev. Lett.
  {\bfseries 106} (2011) 162001},
  \href{http://arxiv.org/abs/1012.3812}{{\ttfamily arXiv:1012.3812 [hep-lat]}}.

\bibitem{Inoue:2010es}
{\bfseries HAL QCD} Collaboration, T.~Inoue, N.~Ishii, S.~Aoki, T.~Doi,
  T.~Hatsuda, Y.~Ikeda, K.~Murano, H.~Nemura, and K.~Sasaki, ``{Bound
  H-dibaryon in Flavor SU(3) Limit of Lattice QCD},''
  \href{http://dx.doi.org/10.1103/PhysRevLett.106.162002}{Phys. Rev. Lett.
  {\bfseries 106} (2011) 162002},
  \href{http://arxiv.org/abs/1012.5928}{{\ttfamily arXiv:1012.5928 [hep-lat]}}.

\bibitem{Luo:2011ar}
Z.-H. Luo, M.~Loan, and Y.~Liu, ``{Search for H dibaryon on the lattice},''
  \href{http://dx.doi.org/10.1103/PhysRevD.84.034502}{Phys. Rev. D {\bfseries
  84} (2011) 034502}, \href{http://arxiv.org/abs/1106.1945}{{\ttfamily
  arXiv:1106.1945 [hep-lat]}}.

\bibitem{Francis:2018qch}
A.~Francis, J.~R. Green, P.~M. Junnarkar, C.~Miao, T.~D. Rae, and H.~Wittig,
  ``{Lattice QCD study of the $H$ dibaryon using hexaquark and two-baryon
  interpolators},'' \href{http://dx.doi.org/10.1103/PhysRevD.99.074505}{Phys.
  Rev. {\bfseries D99} no.~7, (2019) 074505},
\href{http://arxiv.org/abs/1805.03966}{{\ttfamily arXiv:1805.03966 [hep-lat]}}.

\bibitem{Green:2021qol}
J.~R. Green, A.~D. Hanlon, P.~M. Junnarkar, and H.~Wittig, ``{Weakly bound $H$
  dibaryon from SU(3)-flavor-symmetric QCD},''
  \href{http://arxiv.org/abs/2103.01054}{{\ttfamily arXiv:2103.01054
  [hep-lat]}}.

\bibitem{Luscher:1990ck}
M.~Luscher and U.~Wolff, ``{How to Calculate the Elastic Scattering Matrix in
  Two-dimensional Quantum Field Theories by Numerical Simulation},''
\href{http://dx.doi.org/10.1016/0550-3213(90)90540-T}{Nucl. Phys. {\bfseries
  B339} (1990) 222--252}.

\bibitem{PhysRevD.38.298}
M.~Oka, ``Flavor-octet dibaryons in the quark model,''
  \href{http://dx.doi.org/10.1103/PhysRevD.38.298}{Phys. Rev. D {\bfseries 38}
  (Jul, 1988) 298--303}.
  \url{https://link.aps.org/doi/10.1103/PhysRevD.38.298}.

\bibitem{Gal:2015rev}
A.~Gal, ``{Meson assisted dibaryons},''
  \href{http://dx.doi.org/10.5506/APhysPolB.47.471}{Acta Phys. Polon. B
  {\bfseries 47} (2016) 471}, \href{http://arxiv.org/abs/1511.06605}{{\ttfamily
  arXiv:1511.06605 [nucl-th]}}.

\bibitem{STAR:2018uho}
{\bfseries STAR} Collaboration, J.~Adam {\em et~al.}, ``{The Proton-$\Omega$
  correlation function in Au+Au collisions at $\sqrt{s_{NN}}$=200 GeV},''
  \href{http://dx.doi.org/10.1016/j.physletb.2019.01.055}{Phys. Lett. B
  {\bfseries 790} (2019) 490--497},
  \href{http://arxiv.org/abs/1808.02511}{{\ttfamily arXiv:1808.02511
  [hep-ex]}}.

\bibitem{ALICE:2020mfd}
{\bfseries ALICE} Collaboration, S.~Acharya {\em et~al.}, ``{Unveiling the
  strong interaction among hadrons at the LHC},''
  \href{http://dx.doi.org/10.1038/s41586-020-3001-6}{Nature {\bfseries 588}
  (2020) 232--238}, \href{http://arxiv.org/abs/2005.11495}{{\ttfamily
  arXiv:2005.11495 [nucl-ex]}}. [Erratum: Nature 590, E13 (2021)].

\bibitem{PhysRevC.101.015201}
K.~Morita, S.~Gongyo, T.~Hatsuda, T.~Hyodo, Y.~Kamiya, and A.~Ohnishi,
  ``Probing $\mathrm{\ensuremath{\Omega}}\mathrm{\ensuremath{\Omega}}$ and
  $p\mathrm{\ensuremath{\Omega}}$ dibaryons with femtoscopic correlations in
  relativistic heavy-ion collisions,''
  \href{http://dx.doi.org/10.1103/PhysRevC.101.015201}{Phys. Rev. C {\bfseries
  101} (Jan, 2020) 015201}.
  \url{https://link.aps.org/doi/10.1103/PhysRevC.101.015201}.

\bibitem{Meguro:2011nr}
W.~Meguro, Y.-R. Liu, and M.~Oka, ``{Possible $\Lambda_c\Lambda_c$ molecular
  bound state},'' \href{http://dx.doi.org/10.1016/j.physletb.2011.09.088}{Phys.
  Lett. {\bfseries B704} (2011) 547--550},
\href{http://arxiv.org/abs/1105.3693}{{\ttfamily arXiv:1105.3693 [hep-ph]}}.

\bibitem{Vijande:2016nzk}
J.~Vijande, A.~Valcarce, J.~M. Richard, and P.~Sorba, ``{Search for
  doubly-heavy dibaryons in a quark model},''
  \href{http://dx.doi.org/10.1103/PhysRevD.94.034038}{Phys. Rev. D {\bfseries
  94} no.~3, (2016) 034038}, \href{http://arxiv.org/abs/1608.03982}{{\ttfamily
  arXiv:1608.03982 [hep-ph]}}.

\bibitem{Richard:2016eis}
J.-M. Richard, ``{Exotic hadrons: review and perspectives},''
  \href{http://dx.doi.org/10.1007/s00601-016-1159-0}{Few Body Syst. {\bfseries
  57} no.~12, (2016) 1185--1212},
  \href{http://arxiv.org/abs/1606.08593}{{\ttfamily arXiv:1606.08593
  [hep-ph]}}.

\bibitem{Dong:2021bvy}
X.-K. Dong, F.-K. Guo, and B.-S. Zou, ``{A survey of heavy\textendash{}heavy
  hadronic molecules},''
  \href{http://dx.doi.org/10.1088/1572-9494/ac27a2}{Commun. Theor. Phys.
  {\bfseries 73} no.~12, (2021) 125201},
  \href{http://arxiv.org/abs/2108.02673}{{\ttfamily arXiv:2108.02673
  [hep-ph]}}.

\bibitem{Junnarkar:2019equ}
P.~Junnarkar and N.~Mathur, ``{Deuteronlike Heavy Dibaryons from Lattice
  Quantum Chromodynamics},''
  \href{http://dx.doi.org/10.1103/PhysRevLett.123.162003}{Phys. Rev. Lett.
  {\bfseries 123} no.~16, (2019) 162003},
  \href{http://arxiv.org/abs/1906.06054}{{\ttfamily arXiv:1906.06054
  [hep-lat]}}.

\bibitem{Lyu:2021qsh}
Y.~Lyu, H.~Tong, T.~Sugiura, S.~Aoki, T.~Doi, T.~Hatsuda, J.~Meng, and
  T.~Miyamoto, ``{Dibaryon with Highest Charm Number near Unitarity from
  Lattice QCD},'' \href{http://dx.doi.org/10.1103/PhysRevLett.127.072003}{Phys.
  Rev. Lett. {\bfseries 127} no.~7, (2021) 072003},
  \href{http://arxiv.org/abs/2102.00181}{{\ttfamily arXiv:2102.00181
  [hep-lat]}}.

\bibitem{Mathur:2022nez}
N.~Mathur, M.~Padmanath, and D.~Chakraborty, ``{The most beautiful strongly
  bound dibaryon},'' \href{http://arxiv.org/abs/2205.02862}{{\ttfamily
  arXiv:2205.02862 [hep-lat]}}.

\bibitem{BEANE20111}
S.~Beane, W.~Detmold, K.~Orginos, and M.~Savage, ``Nuclear physics from lattice
  qcd,''
  \href{http://dx.doi.org/https://doi.org/10.1016/j.ppnp.2010.08.002}{Progress
  in Particle and Nuclear Physics {\bfseries 66} no.~1, (2011) 1--40}.
  \url{https://www.sciencedirect.com/science/article/pii/S0146641010000530}.

\bibitem{PhysRevD.85.054511}
{\bfseries NPLQCD Collaboration} Collaboration, S.~R. Beane, E.~Chang,
  W.~Detmold, H.~W. Lin, T.~C. Luu, K.~Orginos, A.~Parre\~no, M.~J. Savage,
  A.~Torok, and A.~Walker-Loud, ``Deuteron and exotic two-body bound states
  from lattice qcd,'' \href{http://dx.doi.org/10.1103/PhysRevD.85.054511}{Phys.
  Rev. D {\bfseries 85} (Mar, 2012) 054511}.
  \url{https://link.aps.org/doi/10.1103/PhysRevD.85.054511}.

\bibitem{PhysRevD.73.054503}
{\bfseries NPLQCD Collaboration} Collaboration, S.~R. Beane, P.~F. Bedaque,
  K.~Orginos, and M.~J. Savage, ``$i=2$ $\ensuremath{\pi}\ensuremath{\pi}$
  scattering from fully-dynamical mixed-action lattice qcd,''
  \href{http://dx.doi.org/10.1103/PhysRevD.73.054503}{Phys. Rev. D {\bfseries
  73} (Mar, 2006) 054503}.
  \url{https://link.aps.org/doi/10.1103/PhysRevD.73.054503}.

\bibitem{Ishii:2006ec}
N.~Ishii, S.~Aoki, and T.~Hatsuda, ``{The Nuclear Force from Lattice QCD},''
  \href{http://dx.doi.org/10.1103/PhysRevLett.99.022001}{Phys. Rev. Lett.
  {\bfseries 99} (2007) 022001},
  \href{http://arxiv.org/abs/nucl-th/0611096}{{\ttfamily
  arXiv:nucl-th/0611096}}.

\bibitem{Aoki:2012tk}
{\bfseries HAL QCD} Collaboration, S.~Aoki, T.~Doi, T.~Hatsuda, Y.~Ikeda,
  T.~Inoue, N.~Ishii, K.~Murano, H.~Nemura, and K.~Sasaki, ``{Lattice QCD
  approach to Nuclear Physics},''
  \href{http://dx.doi.org/10.1093/ptep/pts010}{PTEP {\bfseries 2012} (2012)
  01A105}, \href{http://arxiv.org/abs/1206.5088}{{\ttfamily arXiv:1206.5088
  [hep-lat]}}.

\bibitem{PhysRevD.87.034505}
{\bfseries for the Hadron Spectrum Collaboration} Collaboration, J.~J. Dudek,
  R.~G. Edwards, and C.~E. Thomas, ``Energy dependence of the
  $\ensuremath{\rho}$ resonance in $\ensuremath{\pi}\ensuremath{\pi}$ elastic
  scattering from lattice qcd,''
  \href{http://dx.doi.org/10.1103/PhysRevD.87.034505}{Phys. Rev. D {\bfseries
  87} (Feb, 2013) 034505}.
  \url{https://link.aps.org/doi/10.1103/PhysRevD.87.034505}.

\bibitem{PhysRevLett.113.182001}
{\bfseries for the Hadron Spectrum Collaboration} Collaboration, J.~J. Dudek,
  R.~G. Edwards, C.~E. Thomas, and D.~J. Wilson, ``Resonances in coupled
  $\ensuremath{\pi}k\text{\ensuremath{-}}\ensuremath{\eta}k$ scattering from
  quantum chromodynamics,''
  \href{http://dx.doi.org/10.1103/PhysRevLett.113.182001}{Phys. Rev. Lett.
  {\bfseries 113} (Oct, 2014) 182001}.
  \url{https://link.aps.org/doi/10.1103/PhysRevLett.113.182001}.

\bibitem{PhysRevLett.118.022002}
{\bfseries for the Hadron Spectrum Collaboration} Collaboration, R.~A.
  Brice\~no, J.~J. Dudek, R.~G. Edwards, and D.~J. Wilson, ``Isoscalar
  $\ensuremath{\pi}\ensuremath{\pi}$ scattering and the $\ensuremath{\sigma}$
  meson resonance from qcd,''
  \href{http://dx.doi.org/10.1103/PhysRevLett.118.022002}{Phys. Rev. Lett.
  {\bfseries 118} (Jan, 2017) 022002}.
  \url{https://link.aps.org/doi/10.1103/PhysRevLett.118.022002}.

\bibitem{Detmold:2012eu}
W.~Detmold and K.~Orginos, ``{Nuclear correlation functions in lattice QCD},''
  \href{http://dx.doi.org/10.1103/PhysRevD.87.114512}{Phys. Rev. {\bfseries
  D87} no.~11, (2013) 114512},
\href{http://arxiv.org/abs/1207.1452}{{\ttfamily arXiv:1207.1452 [hep-lat]}}.

\bibitem{Doi:2012xd}
T.~Doi and M.~G. Endres, ``{Unified contraction algorithm for multi-baryon
  correlators on the lattice},''
  \href{http://dx.doi.org/10.1016/j.cpc.2012.09.004}{Comput. Phys. Commun.
  {\bfseries 184} (2013) 117},
\href{http://arxiv.org/abs/1205.0585}{{\ttfamily arXiv:1205.0585 [hep-lat]}}.

\bibitem{Humphrey:2022yjc}
N.~Humphrey, W.~Detmold, R.~D. Young, and J.~M. Zanotti, ``{Novel Algorithms
  for Computing Correlation Functions of Nuclei},'' in {\em {38th International
  Symposium on Lattice Field Theory}}.
\newblock 1, 2022.
\newblock \href{http://arxiv.org/abs/2201.04269}{{\ttfamily arXiv:2201.04269
  [hep-lat]}}.

\bibitem{Bicudo:2012qt}
{\bfseries European Twisted Mass} Collaboration, P.~Bicudo and M.~Wagner,
  ``{Lattice QCD signal for a bottom-bottom tetraquark},''
  \href{http://dx.doi.org/10.1103/PhysRevD.87.114511}{Phys. Rev. {\bfseries
  D87} no.~11, (2013) 114511},
\href{http://arxiv.org/abs/1209.6274}{{\ttfamily arXiv:1209.6274 [hep-ph]}}.

\bibitem{Francis:2016hui}
A.~Francis, R.~J. Hudspith, R.~Lewis, and K.~Maltman, ``{Lattice Prediction for
  Deeply Bound Doubly Heavy Tetraquarks},''
  \href{http://dx.doi.org/10.1103/PhysRevLett.118.142001}{Phys. Rev. Lett.
  {\bfseries 118} no.~14, (2017) 142001},
\href{http://arxiv.org/abs/1607.05214}{{\ttfamily arXiv:1607.05214 [hep-lat]}}.

\bibitem{Junnarkar:2018twb}
P.~Junnarkar, N.~Mathur, and M.~Padmanath, ``{Study of doubly heavy tetraquarks
  in Lattice QCD},'' \href{http://dx.doi.org/10.1103/PhysRevD.99.034507}{Phys.
  Rev. {\bfseries D99} no.~3, (2019) 034507},
\href{http://arxiv.org/abs/1810.12285}{{\ttfamily arXiv:1810.12285 [hep-lat]}}.

\bibitem{Leskovec:2019ioa}
L.~Leskovec, S.~Meinel, M.~Pflaumer, and M.~Wagner, ``{Lattice QCD
  investigation of a doubly-bottom $\bar{b} \bar{b} u d$ tetraquark with
  quantum numbers $I(J^P) = 0(1^+)$},''
\href{http://arxiv.org/abs/1904.04197}{{\ttfamily arXiv:1904.04197 [hep-lat]}}.

\bibitem{PhysRevLett.121.202002}
N.~Mathur, M.~Padmanath, and S.~Mondal, ``Precise predictions of charmed-bottom
  hadrons from lattice qcd,''
  \href{http://dx.doi.org/10.1103/PhysRevLett.121.202002}{Phys. Rev. Lett.
  {\bfseries 121} (Nov, 2018) 202002}.
  \url{https://link.aps.org/doi/10.1103/PhysRevLett.121.202002}.

\bibitem{Mathur:2018rwu}
N.~Mathur and M.~Padmanath, ``{Lattice QCD study of doubly-charmed strange
  baryons},'' \href{http://dx.doi.org/10.1103/PhysRevD.99.031501}{Phys. Rev. D
  {\bfseries 99} no.~3, (2019) 031501},
  \href{http://arxiv.org/abs/1807.00174}{{\ttfamily arXiv:1807.00174
  [hep-lat]}}.

\bibitem{Bazavov:2012xda}
{\bfseries MILC} Collaboration, A.~Bazavov {\em et~al.}, ``{Lattice QCD
  ensembles with four flavors of highly improved staggered quarks},''
  \href{http://dx.doi.org/10.1103/PhysRevD.87.054505}{Phys. Rev. {\bfseries
  D87} no.~5, (2013) 054505},
\href{http://arxiv.org/abs/1212.4768}{{\ttfamily arXiv:1212.4768 [hep-lat]}}.

\bibitem{Bazavov:2015yea}
{\bfseries MILC} Collaboration, A.~Bazavov {\em et~al.}, ``{Gradient flow and
  scale setting on MILC HISQ ensembles},''
  \href{http://dx.doi.org/10.1103/PhysRevD.93.094510}{Phys. Rev. {\bfseries
  D93} no.~9, (2016) 094510},
\href{http://arxiv.org/abs/1503.02769}{{\ttfamily arXiv:1503.02769 [hep-lat]}}.

\bibitem{PhysRevD.91.054508}
{\bfseries HPQCD Collaboration} Collaboration, B.~Chakraborty, C.~T.~H. Davies,
  B.~Galloway, P.~Knecht, J.~Koponen, G.~C. Donald, R.~J. Dowdall, G.~P.
  Lepage, and C.~McNeile, ``High-precision quark masses and qcd coupling from
  ${n}_{f}=4$ lattice qcd,''
  \href{http://dx.doi.org/10.1103/PhysRevD.91.054508}{Phys. Rev. D {\bfseries
  91} (Mar, 2015) 054508}.
  \url{https://link.aps.org/doi/10.1103/PhysRevD.91.054508}.

\bibitem{Lepage:1992tx}
G.~P. Lepage, L.~Magnea, C.~Nakhleh, U.~Magnea, and K.~Hornbostel, ``{Improved
  nonrelativistic QCD for heavy quark physics},''
  \href{http://dx.doi.org/10.1103/PhysRevD.46.4052}{Phys. Rev. {\bfseries D46}
  (1992) 4052--4067},
\href{http://arxiv.org/abs/hep-lat/9205007}{{\ttfamily arXiv:hep-lat/9205007
  [hep-lat]}}.

\bibitem{PhysRevD.85.054509}
{\bfseries HPQCD Collaboration} Collaboration, R.~J. Dowdall, B.~Colquhoun,
  J.~O. Daldrop, C.~T.~H. Davies, I.~D. Kendall, E.~Follana, T.~C. Hammant,
  R.~R. Horgan, G.~P. Lepage, C.~J. Monahan, and E.~H. M\"uller, ``The upsilon
  spectrum and the determination of the lattice spacing from lattice qcd
  including charm quarks in the sea,''
  \href{http://dx.doi.org/10.1103/PhysRevD.85.054509}{Phys. Rev. D {\bfseries
  85} (Mar, 2012) 054509}.
  \url{https://link.aps.org/doi/10.1103/PhysRevD.85.054509}.

\bibitem{Wetzorke:1999rt}
I.~Wetzorke, F.~Karsch, and E.~Laermann, ``{Further evidence for an unstable H
  dibaryon?},'' \href{http://dx.doi.org/10.1016/S0920-5632(00)91628-1}{Nucl.
  Phys. Proc. Suppl. {\bfseries 83} (2000) 218--220},
\href{http://arxiv.org/abs/hep-lat/9909037}{{\ttfamily arXiv:hep-lat/9909037
  [hep-lat]}}.

\bibitem{Wetzorke:2002mx}
I.~Wetzorke and F.~Karsch, ``{The H dibaryon on the lattice},''
  \href{http://dx.doi.org/10.1016/S0920-5632(03)01531-7}{Nucl. Phys. Proc.
  Suppl. {\bfseries 119} (2003) 278--280},
  \href{http://arxiv.org/abs/hep-lat/0208029}{{\ttfamily arXiv:hep-lat/0208029
  [hep-lat]}}.
[,278(2002)].

\bibitem{Brown:2014ena}
Z.~S. Brown, W.~Detmold, S.~Meinel, and K.~Orginos, ``{Charmed bottom baryon
  spectroscopy from lattice QCD},''
  \href{http://dx.doi.org/10.1103/PhysRevD.90.094507}{Phys. Rev. {\bfseries
  D90} no.~9, (2014) 094507},
\href{http://arxiv.org/abs/1409.0497}{{\ttfamily arXiv:1409.0497 [hep-lat]}}.

\bibitem{Isgur:1991wq}
N.~Isgur and M.~B. Wise, ``{Spectroscopy with heavy quark symmetry},''
  \href{http://dx.doi.org/10.1103/PhysRevLett.66.1130}{Phys. Rev. Lett.
  {\bfseries 66} (1991) 1130--1133}.

\bibitem{Beane:2003da}
S.~R. Beane, P.~F. Bedaque, A.~Parreno, and M.~J. Savage, ``{Two nucleons on a
  lattice},'' \href{http://dx.doi.org/10.1016/j.physletb.2004.02.007}{Phys.
  Lett. {\bfseries B585} (2004) 106--114},
\href{http://arxiv.org/abs/hep-lat/0312004}{{\ttfamily arXiv:hep-lat/0312004
  [hep-lat]}}.

\bibitem{Davoudi:2011md}
Z.~Davoudi and M.~J. Savage, ``{Improving the Volume Dependence of Two-Body
  Binding Energies Calculated with Lattice QCD},''
  \href{http://dx.doi.org/10.1103/PhysRevD.84.114502}{Phys. Rev. {\bfseries
  D84} (2011) 114502},
\href{http://arxiv.org/abs/1108.5371}{{\ttfamily arXiv:1108.5371 [hep-lat]}}.

\bibitem{Briceno:2013bda}
R.~A. Briceno, Z.~Davoudi, T.~Luu, and M.~J. Savage, ``{Two-nucleon systems in
  a finite volume. II. $^3S_1-^3D_1$ coupled channels and the deuteron},''
  \href{http://dx.doi.org/10.1103/PhysRevD.88.114507}{Phys. Rev. {\bfseries
  D88} no.~11, (2013) 114507},
\href{http://arxiv.org/abs/1309.3556}{{\ttfamily arXiv:1309.3556 [hep-lat]}}.

\bibitem{PhysRevD.90.094507}
Z.~S. Brown, W.~Detmold, S.~Meinel, and K.~Orginos, ``Charmed bottom baryon
  spectroscopy from lattice qcd,''
  \href{http://dx.doi.org/10.1103/PhysRevD.90.094507}{Phys. Rev. D {\bfseries
  90} (Nov, 2014) 094507}.
  \url{https://link.aps.org/doi/10.1103/PhysRevD.90.094507}.

\end{thebibliography}\endgroup


\end{document}